\documentclass[twocolumn]
{revtex4-2}

\usepackage{graphicx}
\usepackage{dcolumn}
\usepackage{bm}
\usepackage{amsmath}
\usepackage{slashed}

\begin{document}

\preprint{AAPM/123-QED}

\title{Higher order electroweak radiative corrections in lepton-proton scattering using covariant approach}

\author{A. Aleksejevs}%
\email{aaleksejevs@mun.ca}
\author{S. Barkanova}%
 \email{sbarkanova@mun.ca}
\author{\underline{M. Ghaffar}}
\email{mghaffar@mun.ca}
 \affiliation{School of Science and the Environment, Grenfell campus, Memorial University of Newfoundland.}

\date{\today}

\begin{abstract}
We perform detailed calculations of electroweak radiative corrections to parity violating lepton scattering with a proton target ($ep$ and $\mu p$) up to quadratic and reducible two-loop level using a covariant approach. Our numerical results are presented at energies relevant for a variety of existing and proposed experimental programs such as $Q_{weak}$, P2, MOLLER, MUSE, and experiments at the EIC. Analysis shows that such corrections at the Next-to-Next-to-Leading Order (NNLO) are quite significant and have to be included in searches of physics beyond the standard model, matching the increasing precision of the future experimental programs at low-energy scales. 
\end{abstract}

\keywords{Suggested keywords}
\maketitle
%

\section{\label{sec:level1}INTRODUCTION:}
The Standard Model (SM) of particle physics has been tested in countless experiments with extraordinary precision. Despite this success, it has limitations in explaining the concept of gravity, dark matter and dark energy, matter-antimatter asymmetry in the Universe, and hierarchies of scale related to the Higgs boson. This opens the door to search for beyond the Standard Model (BSM) physics via direct production of additional particles at high-energy accelerators. However, to date, there is no direct evidence of BSM physics even at the latest 13 TeV energy scale achievable at the Large Hadron Collider (LHC), at CERN. In this scenario, low-energy precision physics plays an essential role in reaching the mass and energy scales not directly accessible at the existing high-energy colliders. These indirect searches are achieved through the precise measurements of well-predicted SM observables, allowing the highly targeted alternative tests for BSM physics.\\
  \indent There are many proposed and working experimental programs that aim to find BSM physics by precisely measuring the physical quantities at low-energy scales. One such observable, widely measured, is the weak charge of the proton $Q^P_{W}$, which defines the strength of proton interaction with other particles via the well-known neutral electroweak force. The electroweak interaction violates parity symmetry first postulated in 1956 \cite{PhysRev.104.254} and experimentally proven in 1957 \cite{PhysRev.105.1413}. Parity violating asymmetry provides a tool to isolate the weak interaction and is given by: 
   \begin{equation}
   A_{PV}=\frac{\sigma_{R}-\sigma_{L}}{\sigma_{R}+\sigma_{L}},
		\label{PV asymmetry formula}
      \end{equation}
	
	where $\sigma_{L,R}$ corresponds to the scattering cross sections in case of an incoming particle beam either left or right polarized.\\
  \indent Recently, many parity-violating electron scattering experiments have been performed and have been proposed to be designed following the latest improvements to precisely measure the SM parameters. One such example is the $Q_{weak}$ experiment \cite{Qweak:2013zxf}, \cite{Qweak:2014xey} at the Thomas Jefferson National Accelerator Facility which used a longitudinally polarized beam of electrons accelerated to $1.16~GeV$ and scattered from a liquid-hydrogen target at a small $4-$momentum transfer squared $-q^2=0.0248~GeV^2$. The most up-to-date value of $A_{PV}$ measured by the $Q_{weak}$ experiment is $-226.5\pm7.3$ (statistical) $\pm5.8$ (systematic) parts per billion (ppb). \cite{qweak.exp}. This asymmetry was then used to determine the $Q^P_{W}$ which was reported to be $0.0719\pm0.0045$, where the uncertainty is one standard deviation.\\
\indent Several proposed experiments are performed using the same technique of polarized electron beam scattering on a liquid-hydrogen target to measure the precise value of $A_{PV}$. One such example is the P2 experiment on the MESA accelerator \cite{P2.exp} that will operate at the small beam energy of $155~MeV$. The objective is to measure $A_{PV}=-39.94$ ppb with a precision of $\Delta A_{PV}=0.56~ ppb~(1.4\%)$ with a small $-q^2=4.5\times 10^{-3}~GeV^2$ \cite{P2.exp}.
  Another one is the MOLLER experiment that will be performed in the Jefferson laboratory at an upgraded beam energy $12~GeV$ and aims to measure the parity violating asymmetry in the scattering of longitudinally polarized electrons off unpolarized electrons with a precision of $0.73~ppb$. That would allow a measurement of the weak charge of the electron at a fractional accuracy of $2.3~\%$ and a determination of the weak mixing angle with uncertainty of $\pm 0.00026~(stat)$ $\pm 0.00013~(syst)$ \cite{Aleksejevs:2010ub}.\\
\indent Electron-Ion Collider (EIC)\cite{EIC.exp} at Brookhaven National Laboratory is another highly anticipated facility that aims to make precision measurements of the constituent quarks and gluons of the proton. It will be the first polarized electron-proton collider where the spins of both electron and proton beams are aligned in a controllable way. The polarized beams will then collide in center-of-mass energies ranging from $\sim 20$ to $\sim 100~GeV$ and upgradable to $\sim 140~GeV$. \\
 \indent In case of muon beam scattering, two important experiments have been proposed. One corresponding to the $\mu p\rightarrow \mu p$ scattering process is a Paul Scherrer Institute (PSI) MUon Scattering Experiment (MUSE) \cite{MUSE:2017dod} that aims to measure and directly compare $ep$ to $\mu p$ elastic scattering at the subpercent level and low momentum transfer. This scattering experiment will be the best test of lepton universality to date and has the potential to demonstrate whether the interactions $\mu p$ and $ep$ are consistent or different. If the discrepancy is real, it should be confirmed with significance of $\approx 5\sigma$.\\ 
 \indent The above-mentioned high-precision parity violating experiments require a new level of accuracy of electroweak radiative corrections which include higher-order effects. These are the processes that are quite more complicated than the actual process but are indistinguishable from it experimentally. A complete set of one-loop-level electroweak radiative corrections in $A_{PV}$ has already been obtained in the case of lepton-nucleon scattering \cite{MUSOLF1992509}, \cite{MUSOLF1990461}, \cite{PhysRevD.43.2956}, \cite{PhysRevD.29.75}, \cite{PhysRevD.70.073002}, \cite{PhysRevD.90.033001} as well as electron-electron scattering \cite{PhysRevD.53.1066}, \cite{Aleksejevs:2010nf}, \cite{Aleksejevs:2012zz}, while some calculations have been performed in order to study quadratic \cite{PhysRevD.85.013007} and two-loop effects in low-energy electroweak measurements \cite{twoloop}, \cite{ALEKSEJEVS20162259}. In 1976, the covariant approach was first introduced by Bardin and Shumeiko \cite{BARDIN-1977} to extract the infrared divergence from the lowest-order bremsstrahlung cross section. This approach has also been used to get explicit expressions in the case of QED radiative corrections up to one-loop level for elastic electron-nucleon scattering \cite{Afanasev-2002}. In this work, we used the covariant approach and calculated the most precise electroweak leptonic tensor up to NNLO (quadratic and reducible two-loop). This leptonic tensor once obtained can be used to calculate the scattering cross-section for any distinguishable hadronic target. In previous calculations of $A_{PV}$, the mass of the electron $(m_{e})$ was treated as a small parameter, but in our case we kept it to account for better precision. We calculate electroweak one-loop (vertex corrections and self-energies), quadratic (squaring of vertex correction and self-energy graphs) and reducible two-loop level $A_{PV}$ using FeynArts \cite{Hahn:2000kx}, FormCalc \cite{HAHN1999153}, FeynCalc \cite{Shtabovenko:2020gxv} and LoopTools \cite{HAHN1999153} as the base languages. Our calculated results are in good agreement with the measured and proposed values of asymmetries of the $Q_{weak}$ and P2 experiments, respectively. The theoretical predictions in this work will be important for the above-mentioned MOLLER, EIC and MUSE experimental programs (either directly or as background studies) searching for physics beyond the Standard Model at the precision frontier. This work is done considering elastic lepton-proton scattering and by using an unpolarized proton. In the future, we would like to consider a hadronic target in an inelastic regime and proton being polarized. In this work, we did not consider the box diagrams; however, one can calculate them using the approaches \cite{PhysRevC.73.055201} and \cite{PhysRevLett.91.142304}.\\
  \indent The paper is organized as follows. The basic notation and introduction to the covariant approach with the tree-level polarized lepton scattering are presented in $section II$. The contributions from higher order (Next-To-Leading Order (NLO) and Next-To-Next-To-Leading Order (NNLO)) corrections in $A_{PV}$ are described in $sec. III$. The electroweak tree-level and one-loop-level hadronic tensor calculations are presented in $sec. IV$. In order to treat infrared (IR) divergence, we use the bremsstrahlung process in soft photon approximation and details are given in $sec. V$. The numerical analysis is presented in $sec. VI$ and conclusions in $sec. VII$. 



\section{COVARIANT APPROACH AND PARITY VIOLATING ASYMMETRY}
\begin{figure}[htb]
		\centering
		\includegraphics[scale=0.15]{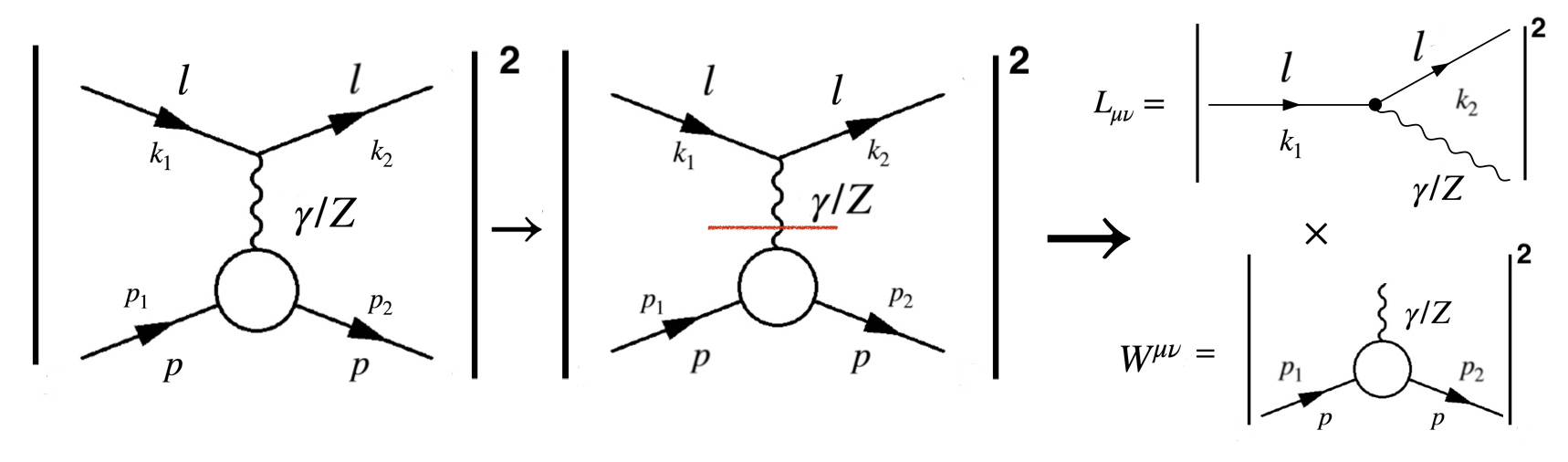}
		\caption{Tree-level Feynman diagrams for lepton-proton $(l-p)$ elastic scattering using a covariant approach. The cross sign represents the contraction between the leptonic ($L_{\mu \nu}$) and hadronic ($W^{\mu \nu}$) currents.}
		\label{fig:leptonicT1}
	\end{figure}

The covariant approach involves a covariant formulation by applying the "cutting rules" such as those of Cutkosky and Landau \cite{Cutkosky:1960sp}. The idea is to relate the imaginary part of a scattering amplitude to physical quantities like cross-sections and is shown in Fig.[\ref{fig:leptonicT1}]. In this way, one gets two parts of the scattering diagram, which are calculated separately and contracted in the end to get the total scattering cross-sections.
 	
The formula for calculating the tree-level parity-violating asymmetry is given in Eq.[\ref{PV asymmetry formula}]. Since $\sigma_R\propto\lvert{\mathcal{M}_R}\rvert^2$ and $\sigma_L\propto\lvert{\mathcal{M}_L}\rvert^2$, one can rewrite $A_{PV}$ in terms of amplitude squared as:
\begin{equation}
A_{PV}=\frac{|\mathcal{M}_{Z}|^2_R-|\mathcal{M}_{Z}|^2_L+2\Re(\mathcal{M}_{\gamma}\mathcal{M}_Z^{\dagger})_R-2\Re(\mathcal{M}_{\gamma}\mathcal{M}_Z^{\dagger})_L}{|\mathcal{M}_{\gamma}|^2_{R+L}+|\mathcal{M}_{Z}|^2_{R+L}+2\Re(\mathcal{M}_{\gamma}\mathcal{M}_Z^{\dagger})_{R+L}},
\label{eq:Apv as amplitude}
\end{equation}
where $(\mathcal{M}_{Z})$ and $(\mathcal{M}_{\gamma})$ are the amplitudes in the case of weak neutral and electromagnetic interactions. The term $\mathcal{M}_{\gamma}\mathcal{M}^{\dagger}_{Z}$ represents the cross term between weak and electromagnetic interactions. Due to the parity-conserving nature of Quantum Electrodynamics (QED), $|\mathcal{M}_{\gamma}|^2_R=|\mathcal{M}_{\gamma}|^2_L$, the numerator of Eq.[\ref{eq:Apv as amplitude}] contains only terms that violate parity.\\
\indent In the Standard Model, considering the case of lepton $(l)$ scattering on a proton $(p)$ target, one can write:
\begin{equation}
l(k_1,s_1)+p(p_1)\rightarrow l(k_2)+p(p_2),
\label{eq:ep scattering eqn}
\end{equation}
where $k_1$ and $~k_2$ are the momenta of the incident and scattered lepton, whereas $p_1$ and $p_2$ are the momenta of the incident and recoiled proton. Considering $l-p$ scattering, the tree-level electromagnetic and weak neutral amplitudes are given as:
\begin{multline}
\mathcal{M}_{\gamma}=[\bar{u}(k_2)(-\dot{\iota}e\gamma_\mu)u^{s(1)}(k_1)]\left(\frac{-\dot{\iota}}{q^2}\right)\\
\times [\bar{u}(p_2)(-\dot{\iota}e\Gamma^\mu_{\gamma p}(q^2)u(p_1)],\\
\\
\mathcal{M}_{Z}=[\bar{u}(k_2)(-\dot{\iota}e(a_V\gamma_\mu+a_{AV}\gamma_\mu \gamma_5))u^{s(1)}(k_1)]\\
\times\left(\frac{-\dot{\iota}}{q^2-m^2_Z}\right)[\bar{u}(p_2)(-\dot{\iota}e\Gamma^\mu_{Zp}(q^2))u(p_1)],
\label{eq:qed and weak amplitudes}
\end{multline}
where the term $s_1$ refers to the polarization of the incident lepton. The coupling of the proton with the photon as a function of momentum transfer squared $q^2$ is written as:
\begin{equation}
\Gamma^{\mu}_{\gamma p}(q^2)=\gamma^\mu F_{1p}^{\gamma}(q^2)+\frac{\dot{\iota}}{2m_p}\sigma^{\mu \nu}q_\nu F_{2p}^{\gamma}(q^2),
\label{eq:qed and weak interactions}
\end{equation}
where $\sigma^{\mu \nu}=\frac{\dot{\iota}}{2}[\gamma^{\mu},\gamma^{\nu}]$. The terms $F_{1p}^{\gamma}(q^2)$ and $F_{2p}^{\gamma}(q^2)$ are the Dirac and Pauli form factors which depend on the momentum transfer squared $q^2=(p_2-p_1)^2=(k_2-k_1)^2=-Q^2$. We can also write them in terms of Sach electric $(G_E)$ and Sach magnetic $(G_M)$ form factors as:	
\begin{equation}
\begin{split}
F_{1p}^{\gamma}(q^2)=\frac{\tau G_M(q^2)+G_E(q^2)}{1+ \tau},\\
\\
F_{2p}^{\gamma}(q^2)=\frac{G_M(q^2)-G_E(q^2)}{1+\tau},
\label{eq:sach form factors}
\end{split}
\end{equation}
where $\tau =-\frac{q^2}{4m^2}$. The Sach form factors are the fourier transform of the electric charge and current density distributions. We use the dipole approximation for $q^2$ dependence of these form factors. The vector $(a_V)$ and axial-vector $(a_{AV})$ coupling strengths in Eq.[\ref{eq:qed and weak amplitudes}] are defined as:
\begin{equation}
a_V=\frac{I_3-2s^2_WQ_f}{2s_Wc_W}.
\label{eq:vector coupling}
\end{equation}
\begin{equation}
a_{AV}=\frac{I_3}{2s_Wc_W}.
\label{eq:pseudo vector coupling}
\end{equation}
Here $s_W\equiv \sin\theta_W$ and $c_W\equiv \cos\theta_W$, with $\theta_W$ being the Weinberg mixing angle. The term $Q_f$ in Eq.[\ref{eq:vector coupling}] is the electric charge, which in the case of $e/\mu$ is $-1$, while $I_3=-\frac{1}{2}$ is the lepton's weak isospin.\\
\indent The coupling of the proton with the $Z-$boson is given as:
    \begin{equation}
	\Gamma^{\mu}_{Zp}(q^2)=\gamma^{\mu}F^Z_{1p}(q^2)+\frac{\dot{\iota}}{2m_p}\sigma^{\mu \nu}q_{\nu}F^Z_{2p}(q^2)+\gamma^{\mu}\gamma_5G^Z_A(q^2),
	\label{eq:electroweak pZ coupling}
	\end{equation}
where $F^Z_{(1,2)p}(q^2)$ \cite{Afanasev:2005ex} are the form factors of the proton neutral weak current and $m_p$ is the mass of the proton. $G^Z_A(q^2)$ stands for the isovector axial form factor which is given by:
\begin{equation*}
	G^Z_A(q^2)=-\tau G_A(q^2)+\Delta s,
	\end{equation*}
where $\tau=+1(-1)$ for protons (neutrons) and $\Delta s$ stands for the strange-quark contribution which we ignored in this work. The isovector axial form factor is normalized at $q^2=0~GeV^2$ to the neutron $\beta-$decay constant as $G_A(0)=+1.267\pm 0.0035$.
The weak form factors are related to the proton and neutron electromagnetic form factors by the expression:
\begin{equation}
	F^Z_{(1,2)p}(q^2)=\frac{1}{4s_Wc_W}\Bigg((1-4s^2_W)F^{\gamma}_{(1,2)p}(q^2)-F^{\gamma}_{(1,2)n}(q^2)\Bigg).
	\label{eq:relation bw proton em and weak neutral ffs}
	\end{equation}

\subsection{Leading Order (LO) electroweak leptonic tensor}\label{tree level leptonic tensor}
\begin{figure}[htb]
	\centering
	\includegraphics[scale=0.29]{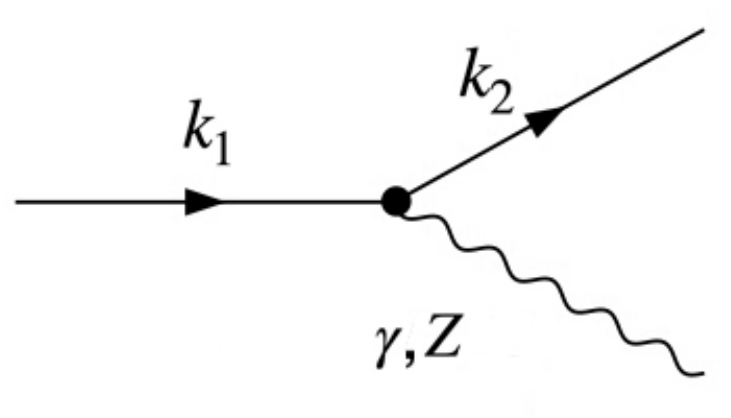}
	\caption{Tree-level electroweak leptonic diagram in case of incoming and outgoing leptons $(l)$ with off-shell $\gamma$, and $Z-$bosons.}
	\label{fig:tree EW leptonic}
\end{figure}

  The LO (tree-level) electroweak polarized lepton $(e,\mu)$ scattering with an unpolarized proton target involves two leptonic tensor diagrams as shown in Fig.[\ref{fig:tree EW leptonic}], having photon $(\gamma)$ and $Z-$boson propagators. Higgs boson interaction is suppressed by its induced propagator and coupling proportional to the mass of lepton, so we ignore it. 

After adding and taking the amplitude squared of the two diagrams with $\gamma$ and $Z-$boson propagators as shown in Fig.[\ref{fig:tree EW leptonic}], 
one gets the tree-level electroweak leptonic tensor as follows:
\begin{multline}
	L_{\mu \nu}^{0}= 4\pi \alpha [l_1g_{\mu \nu}+l_2k_{2\mu}k_{1\nu}+l_3k_{1\mu}k_{2\nu}+\\
l_4\epsilon_{s_1,\mu,\nu,k_1}+l_5\epsilon_{s_1,\mu,\nu,k_2}+l_6\epsilon_{\mu,\nu,k_1,k_2}+\\
 l_7k_{2\mu}s_{1\nu}+l_8k_{2\nu}s_{1\mu}],
		\label{eq:tree EW leptonic tensor full form}
	\end{multline}
where the term like $\epsilon_{\alpha,\mu,\nu,\beta}$ is the Levi-Civita tensor that can be written in terms of helicity reference vector $(s_1=\frac{1}{m_l}(p,0,0,E_1))$ and momentum $(k_2)$ vectors as $\epsilon_{s_1,\mu,\nu,k_2}=s^{\alpha}_1k^{\beta}_2\epsilon_{\alpha,\mu,\nu,\beta}$. The completeness relation we used in the case of polarized leptons is given by:
\begin{equation}
	u_{\beta}^{s_1}(k)\bar{u}^{s_1}_{\beta}(k)=\frac{1}{2}(1+\beta \gamma_5\slashed{s}_1)(\slashed{k}+ m),
	\label{eq:polarized_completeness_relation}
	\end{equation}
where $\beta$ represents the helicity state of the fermions with $+1$ as right-handed and $-1$ as left-handed and $s_1$ is the helicity reference vector. The leptonic structure functions are represented by the terms $l_1-l_8$. In the case of QED, there are only five tensor structures expressed by $g_{\mu \nu},~k_{2\mu}k_{1\nu},~k_{1\mu}k_{2\nu},~\epsilon_{s_1,\mu,\nu,k_1}$ and $~\epsilon_{s_1,\mu,\nu,k_2}$. The analytical details for LO electroweak leptonic tensor structure functions are given in Appendix \ref{leptonic structure functions appendix}.
 
\begin{figure}[htb]
	\centering
	\includegraphics[scale=0.3]{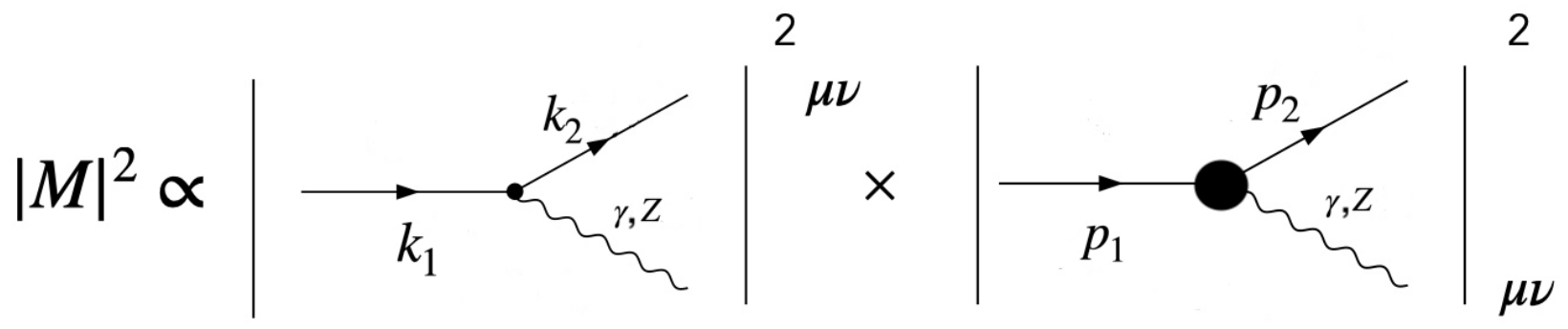}
	\caption{The total amplitude squared in case of tree level electroweak $l-p$ scattering.}
	\label{fig:treeEWtensor}
\end{figure}

 Finally, the total amplitude squared in the case of tree-level electroweak lepton-proton scattering is obtained by contracting the upper leptonic tensor part with the lower hadronic tensor part of the diagram, as shown in Fig.[\ref{fig:treeEWtensor}].
\section{Higher Order Corrections: NLO and NNLO level electroweak leptonic tensor}\label{one-loop and two-loop leptonic tensor}
The higher-order radiative corrections in a scattering process involve the contribution of virtual particles in the form of self-energies, vertex corrections, real photon emissions (bremsstrahlung process) and two-boson exchange (boxes). In this paper, we use the covariant approach to calculate the NLO and NNLO level-boson self-energy and vertex correction graphs, as well as the bremsstrahlung process. Since box diagrams \cite{PhysRevC.73.055201}, \cite{PhysRevLett.91.142304} form a gauge-invariant set, and their calculation with a leptonic tensor approach requires a different treatment, this topic will be discussed in a separate paper. The boson self-energy graphs do not contain infrared (IR) divergences, and we use an on-shell renormalization scheme for the treatment of ultraviolet (UV) divergences. The IR divergence in the vertex correction graphs is treated numerically by using a small parameter for the photon mass. We shall see in sec.\ref{bremsstrahlung} how to get IR finite results analytically by adding bremsstrahlung contributions. 
\subsection{Next-To-Leading Order (NLO) level electroweak leptonic tensor}\label{one-loop level leptonic tensor}
\begin{figure}[htb]
	\centering
	\includegraphics[scale=0.35]{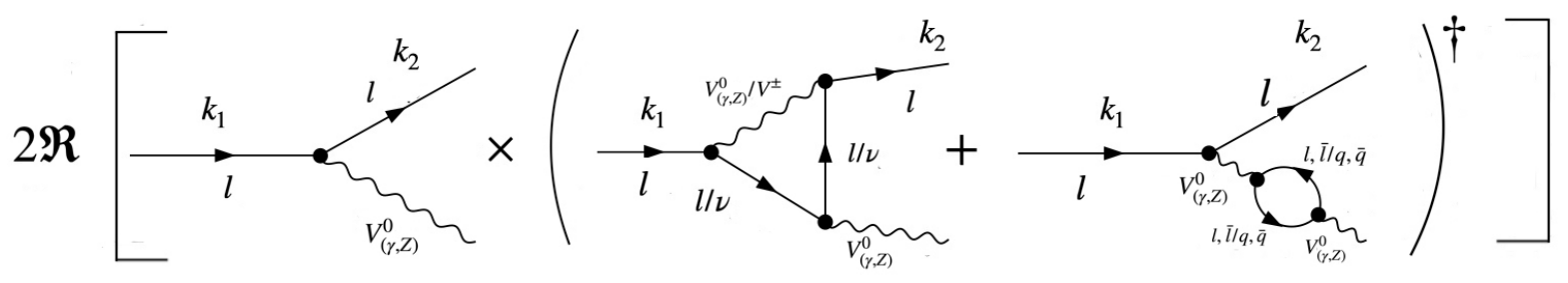}
	\caption{One-loop level electroweak leptonic tensor diagram with an incoming and outgoing lepton. The symbols $V^0$ and $V^{\pm}$ are used for $\gamma$, $Z-$boson and $W^{\pm}$ vector boson  propagators, respectively.}
	\label{fig:EW one loop dig}
\end{figure}

\begin{figure}[htb]
		\centering
		\includegraphics[scale=0.35]{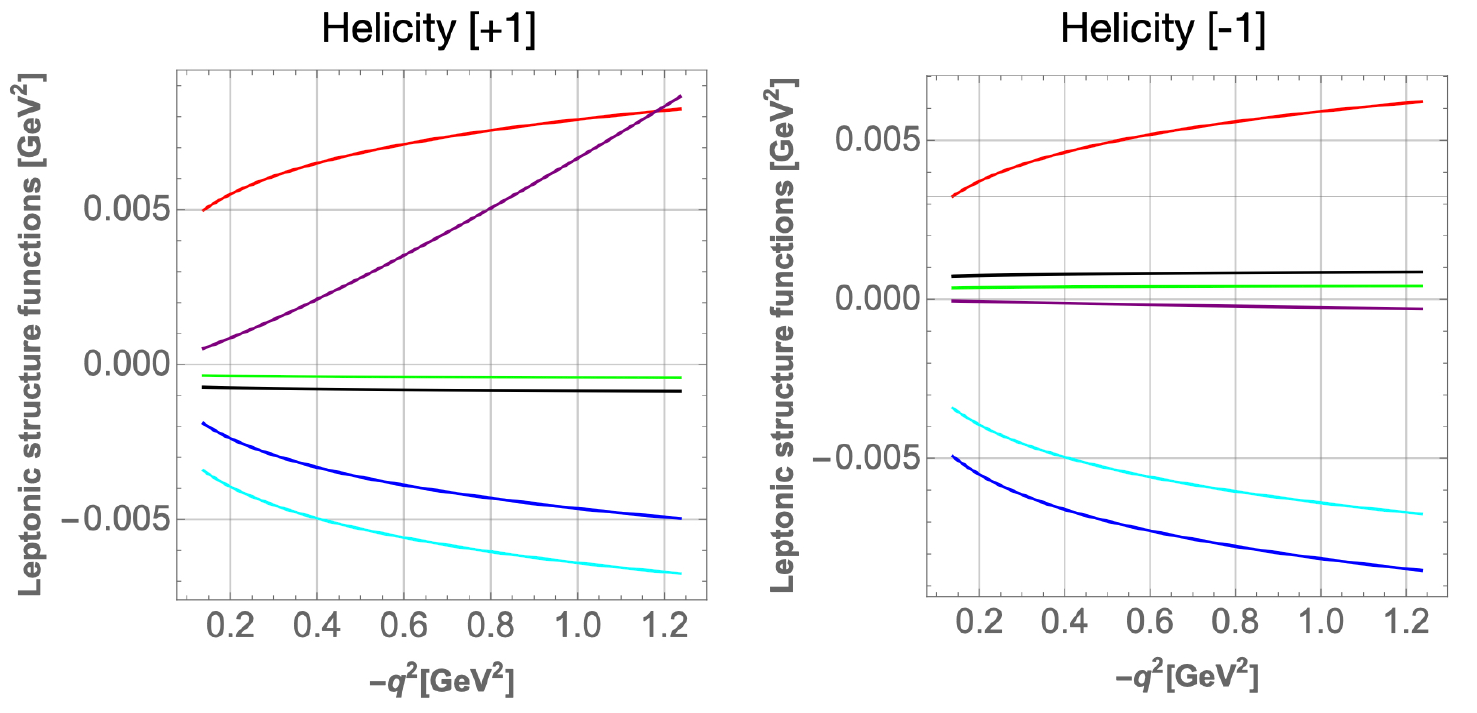}
		\caption{NLO level leptonic tensor structure functions plotted versus $-q^2$ with an incoming polarized lepton having helicity $\pm1$. Here the graphs with colors purple, cyan, blue, red, black and green are for $r_1,~r_2,~r_3,~r_6,~r_{12}$ and $r_{13}$, respectively. The graphs are plotted at $E_{CMS}=20~GeV$.}
		\label{fig:NLO G function plots}
	\end{figure}

The electroweak corrected cross-section at the NLO (one-loop) is of the order of $\alpha^3$ as compared to the LO in the perturbation expansion. It is obtained by multiplying the tree-level diagrams with the sum of the vertex correction and self-energy graphs at the one-loop level, as shown in Fig.[\ref{fig:EW one loop dig}].
This contribution to the total amplitude squared can be expressed as:
\begin{equation}
2\Re[\mathcal{M}_{{LO}}\times
\mathcal{M^\dagger}_{{NLO}}],
\label{eq:NLO full lep amp squared terms}
 \end{equation}
where:
\begin{equation}
\mathcal{M}_{NLO}=\mathcal{\hat{M}}_{SE}+\hat{\mathcal{M}}_{TR}.
\label{eq:se terms defis}
\end{equation}
In Eq.[\ref{eq:se terms defis}], the terms $\mathcal{\hat{M}}_{(SE,TR)}$ correspond to the renormalized photon and $Z-$boson amplitudes in the on-shell renormalization scheme \cite{Hollik:1988ii}, \cite{Denner:1991kt} whereas the subscripts SE and TR represent the self-energy and vertex correction terms, respectively. The vertex correction graphs are infrared divergent that arises due to the massless photon in the vertex loop integral. We will treat this divergence in $sec. V$. \\ 
\indent Using Eq.[\ref{eq:NLO full lep amp squared terms}], we calculate the one-loop electroweak leptonic tensor given by:
\begin{multline}
L_{\mu \nu}^{NLO}=r_1g_{\mu \nu}+r_2k_{2\mu}k_{1\nu}+r_3k_{1\mu} k_{2\nu}+r_4\epsilon_{s_1,\mu,\nu,k_1}+\\
r_5\epsilon_{s_1,\mu,\nu,k_2}+r_6\epsilon_{\mu,\nu,k_1,k_2}+r_7k_{2\mu}s_{1\nu}+r_8k_{2\nu}s_{1\mu}+r_9k_{2\mu}k_{2\nu}+\\
r_{10}\epsilon_{s_1,\mu,k_1,k_2}k_{2,\nu}+r_{11}\epsilon_{s_1,\nu,k_1,k_2}~k_{2\mu}+r_{12}k_{1\nu}k_{1\mu}+\\
r_{13}\epsilon_{\mu,\nu,k_2,k_1}+r_{14}s_{1\nu}k_{1\mu}+r_{15}s_{1\mu}k_{1\nu}+r_{16}\epsilon_{s_1,\mu,k_1,k_2}k_{1\nu}+\\
r_{17}\epsilon_{s_1,\mu,k_2,k_1}k_{2\nu}+r_{18}\epsilon_{s_1,\nu,k_2,k_1}k_{1\mu}+r_{19}\epsilon_{s_1,\nu,k_2,k_1}k_{2\mu}.
	\label{eq:total one loop EW leptonic tensor}
\end{multline}
There are in total nineteen leptonic tensor structure functions obtained by the products of photon couples to lepton $\times$ photon couples to lepton, $Z-$boson couples to lepton $\times$ $Z-$boson couples to lepton, and photon couples to lepton $\times$ $Z-$boson couples to lepton, respectively.\\
\indent In Eq.[\ref{eq:total one loop EW leptonic tensor}], $r_{1-19}$ are the one-loop level leptonic structure functions. These structure functions depend on momentum transfer squared $(-q^2=Q^2)$ and are written in terms of the Passarino-Veltman functions. We use the LoopTools package to calculate these functions numerically. We plot these structure functions as a function of $-q^2$ as shown in Fig.[\ref{fig:NLO G function plots}]. By applying the sensitivity study, it was noticed that the significant contribution to parity violating asymmetry comes from only six structure functions ($r_1,~r_2,~r_3,~r_6,~r_{12}$ and $r_{13}$) which are plotted in Fig.[\ref{fig:NLO G function plots}]. However, the total $A_{PV}$ is calculated by considering all of the above mentioned nineteen structure functions. The structure function $r_{1}$ has the units of $GeV^2$. Hence, in order to make units consistent for all the structure functions, we multiply and divide $r_2,~r_3,~r_6,~r_{12}$ and $r_{13}$ by a scaling parameter $\delta^2=1~GeV^2$.  
     
\begin{figure}[htb]
	\centering
	\includegraphics[scale=0.25]{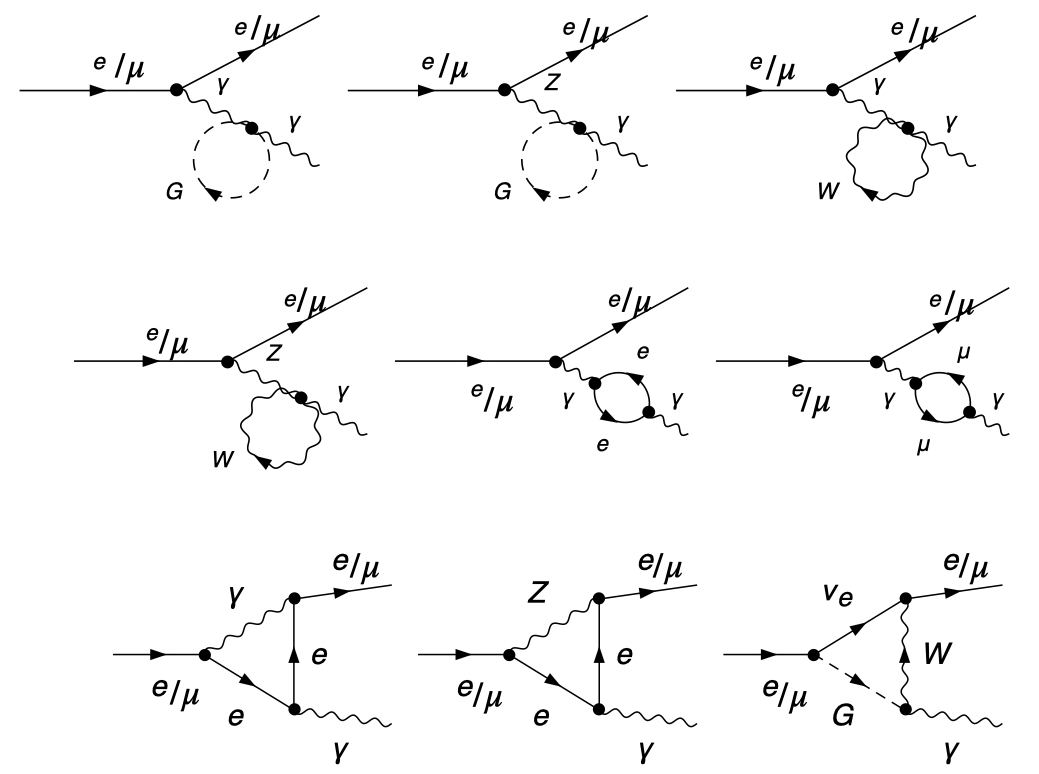}
	\caption{Examples of self-energy and vertex correction diagrams in case of electroweak leptonic tensor at one-loop level. In total there are $307$ graphs.}
	\label{fig:EW one loop exp}
\end{figure}

 There are in total 307 self-energy and vertex correction graphs in the case of electroweak leptonic tensor at the one-loop containing all SM particles inside the loop. A snapshot of such graphs generated in FeynArts is shown in Fig.[\ref{fig:EW one loop exp}].
\subsection{Next-To-Next-To-Leading Order (NNLO) electroweak leptonic tensor}\label{quad level leptonic tensor}
\begin{figure}[htb]
	\centering
	\includegraphics[scale=0.35]{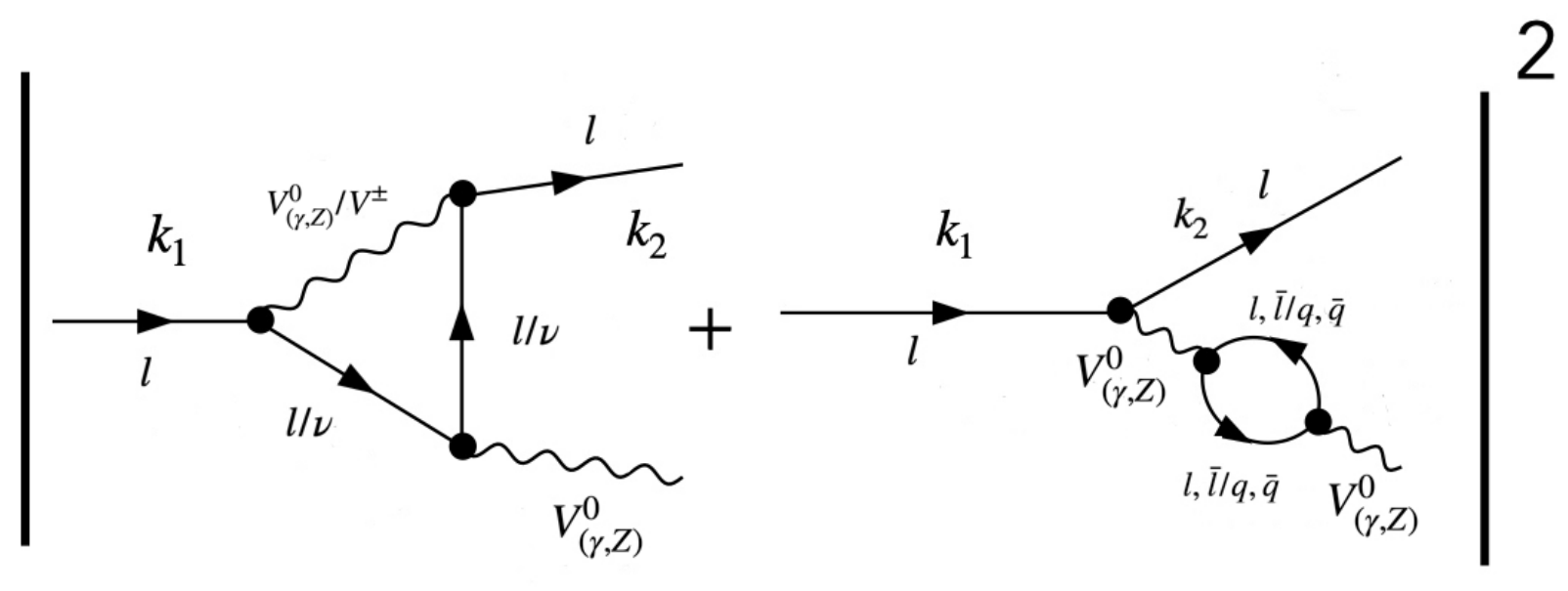}
	\caption{Quadratic level electroweak graphs obtained by squaring the sum of one-loop level self-energy and vertex correction. Here $V^0_{(\gamma,Z)}$ could be either photon or $Z-$boson and $V^{\pm}$ is for $W^{\pm}-$boson.}
	\label{fig:EW quadratic fig}
\end{figure}

The squared amplitude related to the calculations of NNLO electroweak leptonic tensor can be written as:
\begin{equation}
|\mathcal{M}_{NNLO}|^2=|\mathcal{M}_{1l}|^2+2\Re{\mathcal{M}_{LO}(\mathcal{M}^{\dagger}_{2lr}+\mathcal{M}^{\dagger}_{2li}})
\end{equation}
where $\mathcal{M}_{1l}$, $\mathcal{M}_{2lr}$ and $\mathcal{M}_{2li}$ are the one-loop, two-loop reducible as well as two-loop irreducible amplitudes, respectively. At NNLO, the electroweak corrections to the cross-section are of the order of $\alpha^4$. The quadratic electroweak leptonic tensor is obtained by squaring the sum of self-energy and vertex correction graphs at one-loop level, shown in Fig.[\ref{fig:EW quadratic fig}].

\subsubsection{Quadratic NNLO level electroweak leptonic tensor}\label{quad nnlo level leptonic tensor}
\begin{figure}[htb]
		\centering
		\includegraphics[scale=0.35]{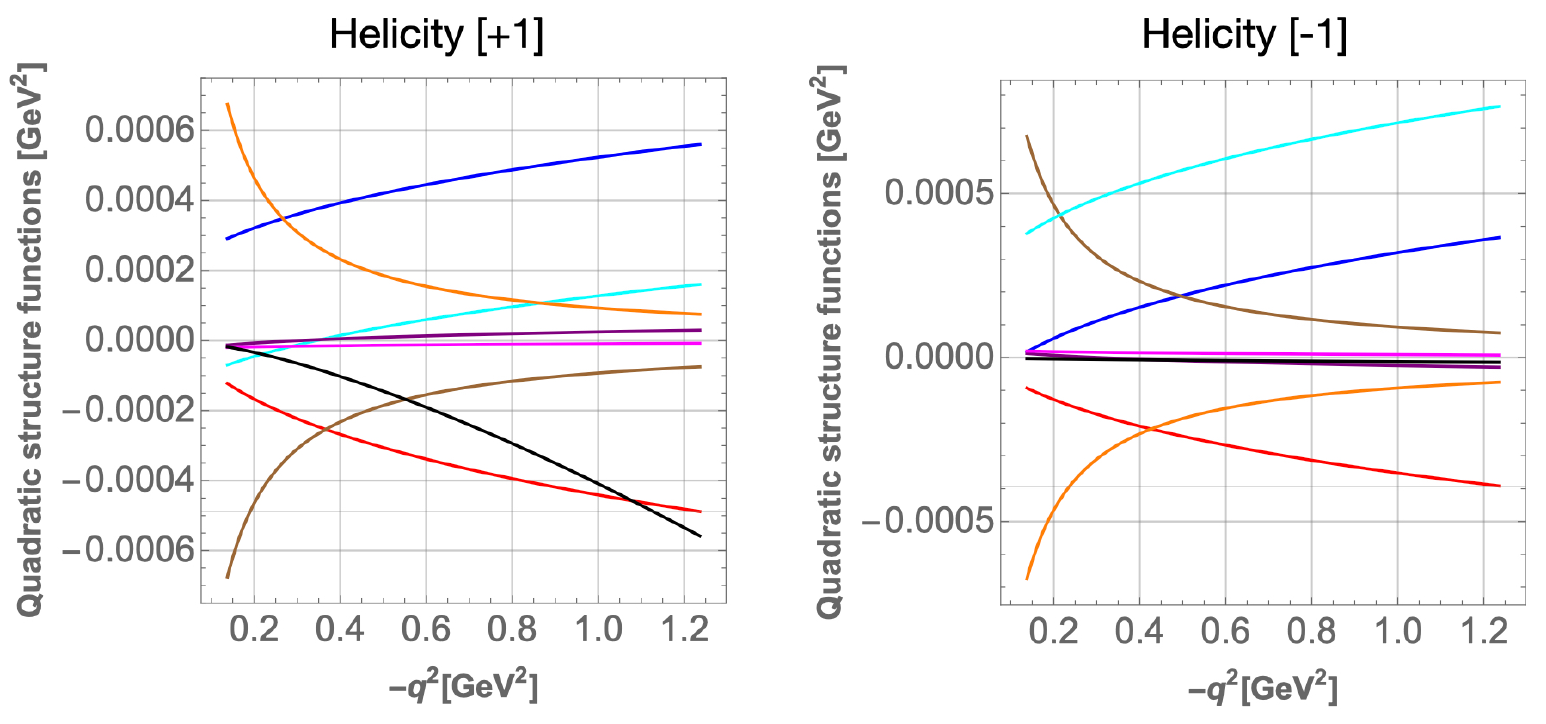}
		\caption{Quadratic level leptonic tensor structure functions plotted versus $-q^2$ with helicities $\pm1$. Here the graphs with colors black, cyan, blue, red, purple, magenta, orange and brown are for $n_1,~n_2,~n_3,~n_6,~n_{12},~n_{13},~n_{14}$ and $n_{15}$, respectively. The graphs are plotted at $E_{CMS}=20~GeV$.}
		\label{fig:Qud G function plots}
	\end{figure}

The electroweak quadratic leptonic tensor $(L^{QD}_{\mu \nu})$ obtained in this way has the following form:
\begin{multline}
	L_{\mu \nu}^{QD}=n_1g_{\mu \nu}+n_2k_{2\mu}k_{1\nu}+n_3k_{1\mu} k_{2\nu}+n_4\epsilon_{s_1,\mu,\nu,k_1}+\\
n_5\epsilon_{s_1,\mu,\nu,k_2}+n_6\epsilon_{\mu,\nu,k_1,k_2}+n_7k_{2\mu}s_{1\nu}+n_8k_{2\nu}s_{1\mu}+n_9k_{2\mu}k_{2\nu}+\\
n_{10}\epsilon_{s_1,\mu,k_1,k_2}k_{2,\nu}+n_{11}\epsilon_{s_1,\nu,k_1,k_2}~k_{2\mu}+n_{12}k_{1\nu}k_{1\mu}+\\
n_{13}\epsilon_{\mu,\nu,k_2,k_1}+n_{14}s_{1\nu}k_{1\mu}+n_{15}s_{1\mu}k_{1\nu}+n_{16}\epsilon_{s_1,\mu,k_1,k_2}k_{1\nu}+\\
n_{17}\epsilon_{s_1,\mu,k_2,k_1}k_{2\nu}+n_{18}\epsilon_{s_1,\nu,k_2,k_1}k_{1\mu}+n_{19}\epsilon_{s_1,\nu,k_2,k_1}k_{2\mu}+\\
n_{20}\epsilon_{s_1,\mu,k_2,k_1}k_{1\nu}+n_{21}\epsilon_{s_1,\nu,k_1,k_2}k_{1\mu}.
		\label{eq:quadratic EW leptonic tensor}
	\end{multline}
where $n_{1-21}$ are the quadratic leptonic structure functions and are calculated using the FormCalc and LoopTools packages. Among these structure functions, only eight ($n_1,~n_2,~n_3,~n_6,~n_{12},~n_{13},~n_{14},~n_{15})$ have a significant contribution to $A_{PV}$ which are plotted in Fig.[\ref{fig:Qud G function plots}] as a function of momentum transfer squared. All these structure functions have different units. In order to keep units consistent ($GeV^2$), we multiply $n_{14},~n_{15}$ by a scaling parameter $\delta=1~GeV$ and $n_2,~n_3,~n_6,~n_{12}$ and $n_{13}$ by $\delta^2=1~GeV^2$. 
\subsubsection{Reducible two-loop level electroweak leptonic tensor}\label{two-loop level leptonic tensor}
\begin{figure*}[htb]
	\centering
	\includegraphics[scale=0.5]{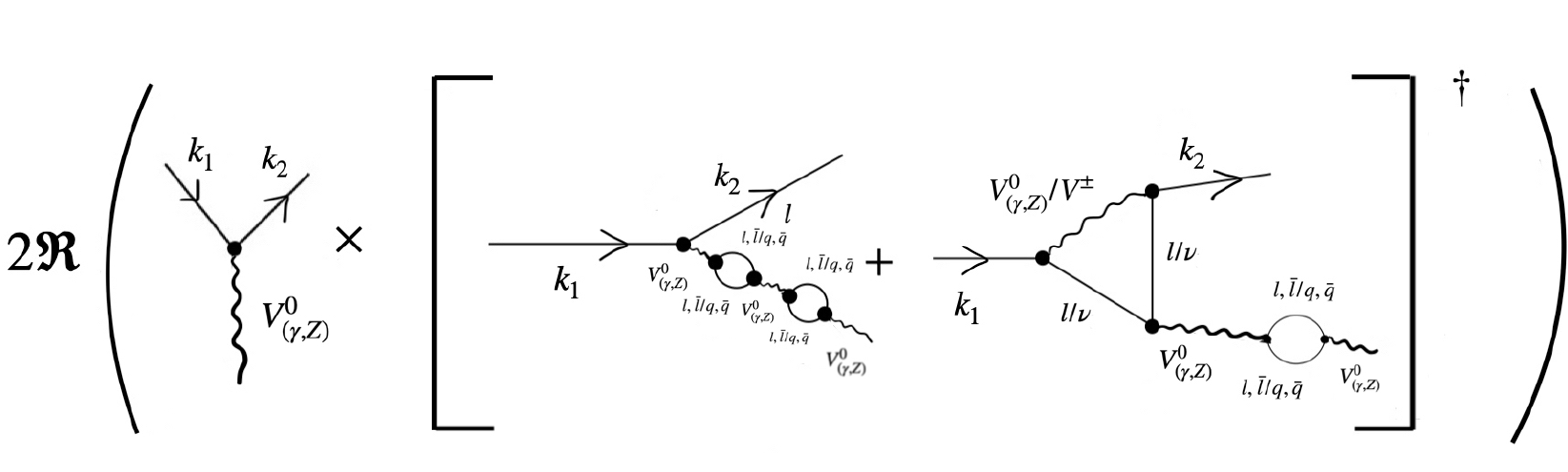}
	\caption{Example of contributions coming in reducible two-loop level electroweak leptonic tensor.}
	\label{fig:two loop fig}
\end{figure*}

\begin{figure*}[htb]
	\centering
	\includegraphics[scale=0.5]{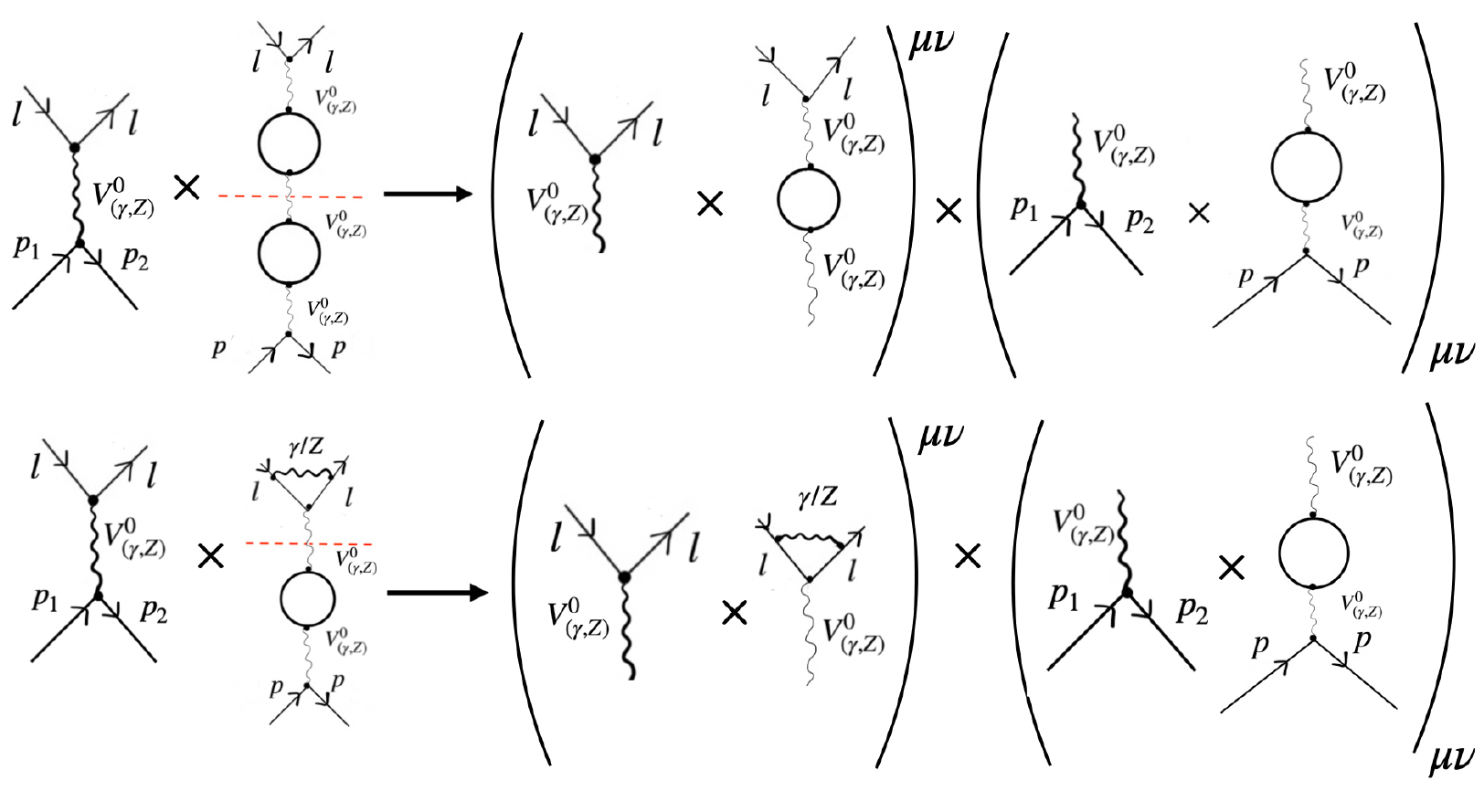}
	\caption{Splitting of the reducible two-loop level diagram into a single loop one at each end of the leptonic and hadronic sides.}
	\label{fig:two loop sep fig}
\end{figure*}

Another way of obtaining the electroweak radiative corrections which are of the order of $\alpha^4$ is by multiplying the tree-level diagram with the reducible two-loop graphs. Such graphs include a double self-energy and a vertex correction attached with a self-energy diagrams, as shown in Fig.[\ref{fig:two loop fig}].
\\ The leptonic tensor obtained in this case has the same form as given in Eq.[\ref{eq:total one loop EW leptonic tensor}], having a total of nineteen lepton structure functions for each $\gamma$, $Z$ and $\gamma-Z$ interaction case. There are two ways of calculating such reducible two-loop level diagrams in Mathematica; one by using an effective propagator approach and the other by considering one of the loops at the upper part (leptonic side) and the other one at the lower part (hadronic side) of the diagram. The final squared amplitude is then obtained by contracting both the upper and lower parts of the diagram. \\
\indent  In this work, we use the second approach to calculate the total squared amplitude of the reducible two-loop level diagram by splitting the two loops at each end of the leptonic-hadronic sides, as shown in Fig.[\ref{fig:two loop sep fig}]. The one-loop level self-energy and vertex correction parts can then be calculated using FeynArts, FormCalc and LoopTools packages. 
\section{ELECTROWEAK HADRONIC TENSOR}\label{hadronic tensor}
In this section we calculate tree-level and one-loop level electroweak hadronic tensor for unpolarized proton target.
\subsection{Tree-level electroweak hadronic tensor}\label{tree hadronic tensor}
\begin{figure}[htb]
		\centering
		\includegraphics[scale=0.5]{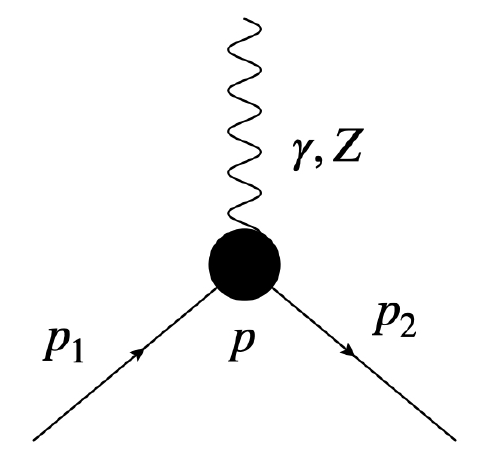}
		\caption{Tree-level electroweak hadronic diagrams with off-shell $\gamma$ and $Z-$bosons.}
		\label{fig:EW had tensor}
	\end{figure}

	The tree-level electroweak hadronic tensor in the case of unpolarized incoming and outgoing protons consists of two diagrams with $\gamma$ and $Z$-boson propagators as shown in Fig.[\ref{fig:EW had tensor}]. Just like in the case of leptonic tensor, the $H$ propagator contribution is ignored here as well.  

\begin{figure}[htb]
		\centering
		\includegraphics[scale=0.45]{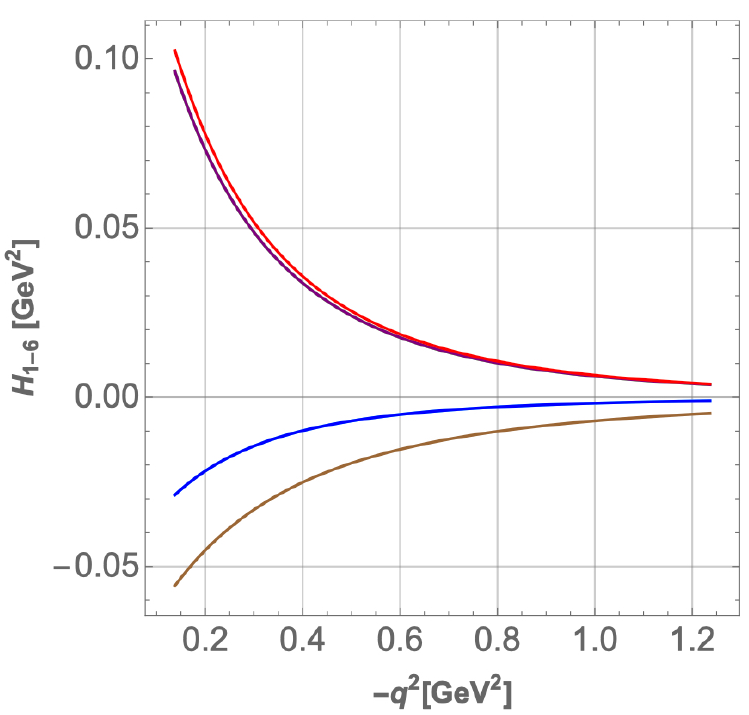}
		\caption{Tree level electroweak hadronic tensor structure functions plotted versus momentum transfer squared $(-q^2)$. Here the graphs with colors red, blue, purple, cyan, green and brown are for $H_1,~H_2,~H_3,~H_4,~H_5$ and $H_6$, respectively. The structure functions $H_4=H_2$ and $H_5=H_3$. The graphs are plotted at $E_{CMS}=20~GeV$.}
		\label{fig:LO had fns plots}
	\end{figure}

 	Using the couplings given in Eqs.[\ref{eq:qed and weak interactions}] and [\ref{eq:electroweak pZ coupling}], we can calculate the Feynman amplitudes of the diagrams shown in Fig.[\ref{fig:EW had tensor}]. The tree-level unpolarized hadronic tensor is calculated as:
 	\begin{multline}
				W^{\mu \nu,VV}_{0}=H^{VV}_{1}~g^{\mu \nu}+H^{VV}_{2}~p^\mu_2p^\nu_1+H^{VV}_{3}~p^\mu_1p^\nu_2+\\
    H^{VV}_{4}~p^\mu_1p^\nu_1+H^{VV}_{5}~p^\mu_2p^\nu_2+H^{VV}_{6}~\epsilon^{\mu, \nu, p_1, p_2}.
			\label{eq:had tensor gg zz gz}
		\end{multline}
   In Eq.[\ref{eq:had tensor gg zz gz}], $VV$ represents $\gamma \gamma$, $ZZ$ or $\gamma Z$ interference. The terms $W^{\mu \nu,\gamma \gamma}_{0}$, $W^{\mu \nu,ZZ}_{0}$ and $W^{\mu \nu,\gamma Z}_{0}$ are obtained from the products of photon couples to proton $\times$ photon couples to proton, $Z-$boson couples to proton $\times$ $Z-$boson couples to proton, and photon couples to proton $\times$ $Z-$boson couples to proton, respectively. The terms $H^{VV}_{1-6}$ represent the hadronic structure functions corresponding to the hadronic tensors $g^{\mu \nu}$,~$p^{\mu}_2p^{\nu}_1$,~$p^{\mu}_1p^{\nu}_2$,~$p^{\mu}_1p^{\nu}_1$,~$p^{\mu}_2p^{\nu}_2$ and $\epsilon^{\mu, \nu, p_1, p_2}$, respectively. These structure functions are written in terms of QED and weak proton form factors which are functions of momentum transfer. These structure functions are shown in Fig.[\ref{fig:LO had fns plots}]. Here we apply the same approach as in lepton structure functions to keep units consistent in terms of $GeV^2$. The analytical details of these structure functions are given in Appendix \ref{hadronic structure functions appendix}.
   \subsection{One-loop level electroweak hadronic tensor}\label{one-loop hadronic tensor}
\begin{figure}[htb]
		\centering
		\includegraphics[scale=0.45]{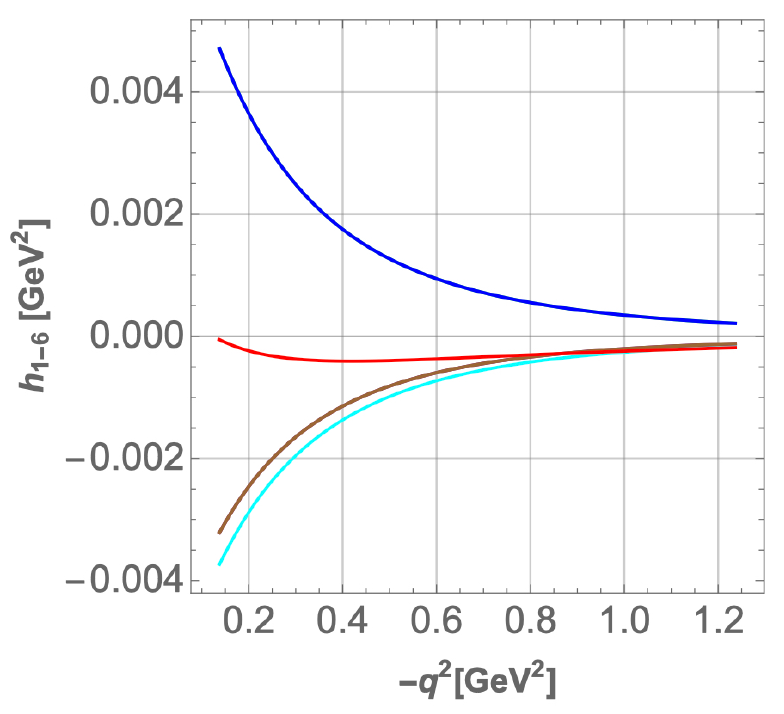}
		\caption{NLO level electroweak hadronic tensor structure functions plotted versus momentum transfer squared $(-q^2)$. Here the graphs with colors red, purple, brown, green, blue and cyan are for $h_1,~h_2,~h_3,~h_4,~h_5$ and $h_6$, respectively. The structure functions $h_4=h_5$ and $h_2=h_3$. The graphs are plotted at $E_{CMS}=20~GeV$.}
		\label{fig:NLO had fns1 plots}
	\end{figure}

   As discussed in subsec.~\ref{two-loop level leptonic tensor}, in order to calculate the reducible two-loop level graphs, we split one of the self-energy to the lower (hadronic) part of the diagram as shown in Fig.[\ref{fig:two loop sep fig}]. After the splitting of reducible two-loop graph into one-loop self-energy, we calculate the electroweak hadronic self-energy part in the same way as the electroweak leptonic self-energy, except now we consider the unpolarized protons. The vertex correction to the hadronic current is not considered in this case. We replace the $\gamma-e$ and $Z-e$ couplings with $\gamma-p$ and $Z-p$ ones as given in Eq.[\ref{eq:electroweak pZ coupling}]. The structure of a one-loop level unpolarized hadronic tensor obtained in this way is given by: 
   \begin{multline}
W^{\mu \nu,VV}_{NLO}=h^{VV}_{1}~g^{\mu \nu}+h^{VV}_{2}~p^\mu_2p^\nu_1+h^{VV}_{3}~p^\mu_1p^\nu_2+\\
h^{VV}_{4}~p^\mu_1p^\nu_1+h^{VV}_{5}~p^\mu_2p^\nu_2+h^{VV}_{6}~\epsilon^{\mu, \nu, p_1, p_2}.
	\label{eq:one loop EW hadronic tensor}
\end{multline}
There are in total \textbf{six} one-loop level unpolarized hadronic tensor structure functions for each photon $(\gamma)$, $Z$ or $\gamma-Z$ interference case. In Eq.[\ref{eq:one loop EW hadronic tensor}], $h^{VV}_{1-6}$ are the hadronic structure functions at the one-loop level. These structure functions depend on the momentum transfer squared $q^2=(p_2-p_1)^2$ between the incoming and outgoing protons and are given in terms of the Passarino-Veltman integral functions. We plot these structure functions as shown in Fig.[\ref{fig:NLO had fns1 plots}]. The units of all these structure functions are made consistent by using a scaling parameter $\delta^2=1~GeV^2$.

The relation between tree and one-loop level hadronic tensor structure functions is given in terms of truncated transverse self-energy $(\hat{\Sigma}^T_{VV})$:
\begin{equation}
\begin{split}
\hat{\Pi}^{VV}_{\mu \nu}=\dot{\iota}\Bigg(g_{\mu \nu}-\frac{k_{\mu}k_{\nu}}{k^2}\Bigg)\hat{\Sigma}^T_{VV}+\frac{k_{\mu}k_{\nu}}{k^2}\hat{\Sigma}^L_{VV},\\
\\
h_i=\frac{H_i}{q^2}\Bigg(\hat{\Sigma}^T_{\gamma \gamma}+2\hat{\Sigma}^T_{\gamma Z}+\hat{\Sigma}^T_{ZZ}\Bigg),
\label{eq:relation bw tree nlo had}
\end{split}
\end{equation}
where $\hat{\Pi}^{VV}_{\mu \nu}$ is the $V-V$ mixing tensor and the terms $\hat{\Sigma}^T_{VV}$ and $\hat{\Sigma}^L_{VV}$ are the transverse and longitudinal parts of the truncated self-energy. The longitudinal part does not contribute in the the cross-section.\\
\indent Once the electroweak (tree and NLO) hadronic tensors are obtained, we contract them with the electroweak leptonic tensors as calculated in subsec.~\ref{tree level leptonic tensor} and sec.~\ref{one-loop and two-loop leptonic tensor}. This contraction gives the total amplitude squared $(|\mathcal{M}|^2)$ which is used to calculate the parity-violating asymmetry given by Eq.[\ref{eq:Apv as amplitude}].
\section{Soft Photon Bremsstrahlung (SPB)}\label{bremsstrahlung}
 The infrared (IR) divergence appears in the vertex correction graphs and fermion self-energies due to lower integral limit of the photon loop momentum $k_{\gamma}\rightarrow 0~GeV$. This IR divergence is regularized numerically by giving a small mass $(\lambda)$ to the photon. Photon mass parameter $\lambda$ is unphysical and exactly cancels out with the lower integration limit of the soft photon bremsstrahlung contribution.
 \subsection{Infrared divergence treatment at NLO level}\label{IR at NLO}
\begin{figure}[htb]
		\centering
		\includegraphics[scale=0.5]{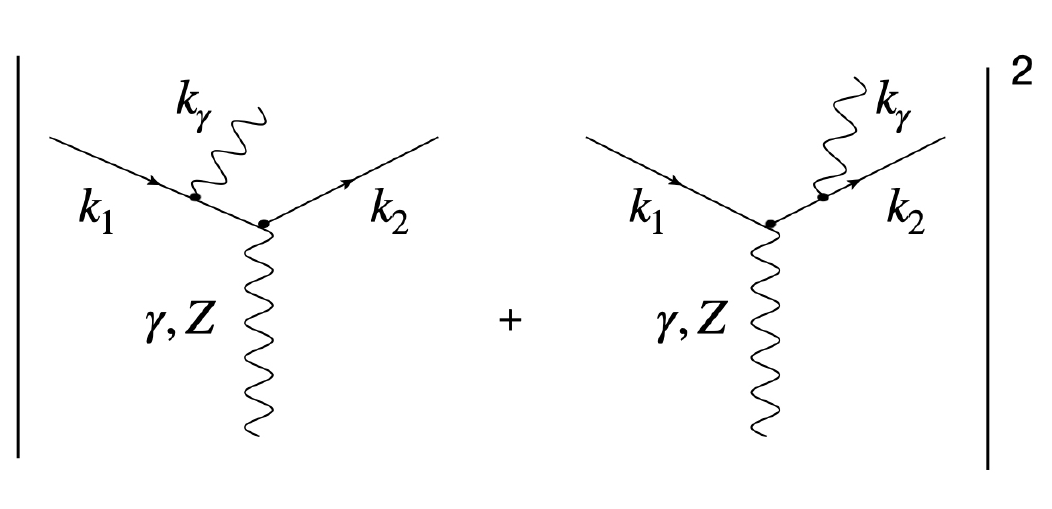}
		\caption{One-loop level electroweak bremsstrahlung process with off-shell $\gamma$ and $Z-$bosons.}
		\label{fig:oneloopfullbremdig.jpeg}
	\end{figure}

We calculate the electroweak bremsstrahlung contribution as shown in Fig.[\ref{fig:oneloopfullbremdig.jpeg}].
The amplitude for the initial state bremsstrahlung $(\mathcal{M}_{in,SPB})^{\beta}$ can be written as:
		\begin{multline}
(\mathcal{M}^{\gamma}_{in,SPB})^{\beta}=\Bigg[\bar{u}(k_2)(\dot{\iota}e\gamma ^\beta)\\
\times \left(\frac{m_l+\slashed{k_1}-\slashed{k}_{\gamma}}{(k_1-k_{\gamma})^2-m^2_l}\right)(\dot{\iota}e\gamma ^\alpha)u(k_1)\Bigg] \epsilon^*_{\alpha}(k_{\gamma}),\\
\\
(\mathcal{M}^{Z}_{in,SPB})^{\beta}=\Bigg[\bar{u}(k_2)(\dot{\iota}e(a_{v}\gamma ^{\beta}+a_p\gamma^{\beta}\gamma_5)\\
\times \left(\frac{m_l+\slashed{k_1}-\slashed{k}_{\gamma}}{(k_1-k_{\gamma})^2-m^2_l}\right)(\dot{\iota}e\gamma ^\alpha)u(k_1)\Bigg] \epsilon^*_{\alpha}(k_{\gamma}).
			\label{eq:incomingG}
		\end{multline}
 The $4-$momentum of the radiated soft photon is given by $k_{\gamma}$ and polarization vector of the emitted soft photon is represented by $\epsilon^*_{\alpha}(k_{\gamma})$.
	
\begin{figure}[!htb]
		\centering
		\includegraphics[scale=0.5]{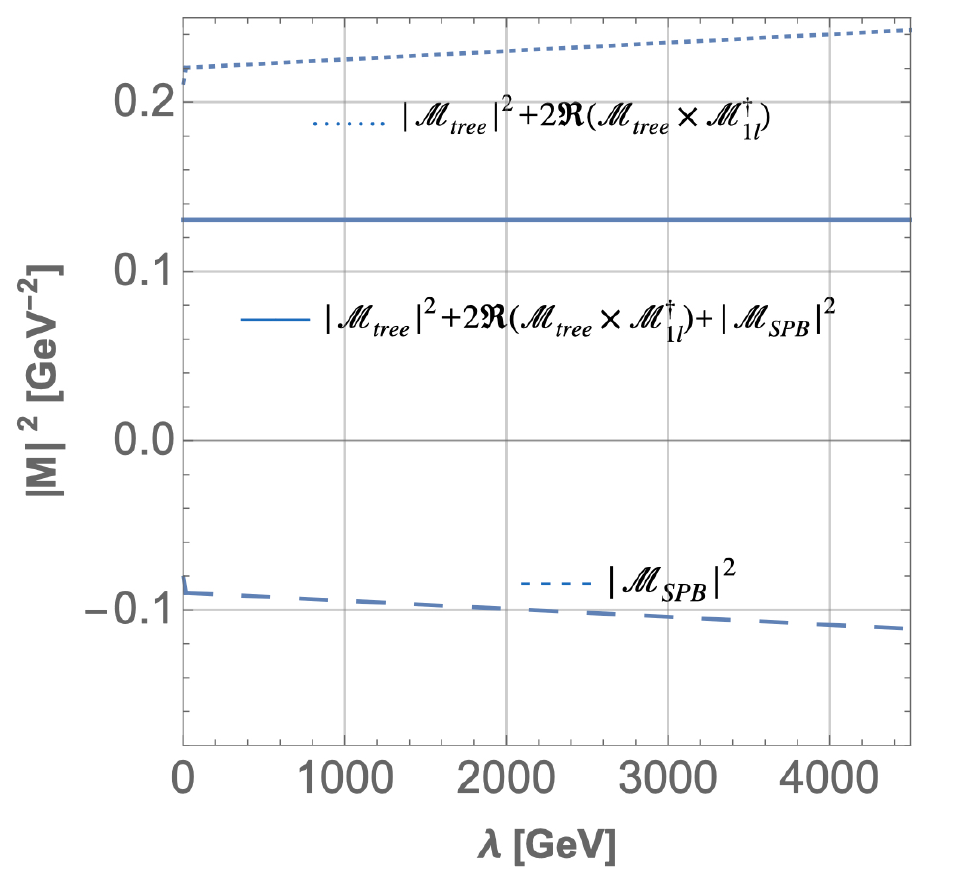}
		\caption{Test for IR-divergence cancellation using MUSE $(\mu-p)$ kinematics at $\theta_{lab}=35^0$ and beam energy $E_{lab}=0.235 ~GeV$. Dashed line is for the SPB amplitude squared, dotted line is the up to NLO level corrections and solid line is the sum of the two mentioned contributions.}
		\label{fig:musebrem.pdf}
	\end{figure}

    Since the energy of the real photon emitted in the SPB process is so small, we can ignore it in the numerator of Eq.[\ref{eq:incomingG}]. Similarly, the denominator of Eq.[\ref{eq:incomingG}] can also be simplified as $-2(k_1\cdot k_{\gamma})$, since $k_1^2=m^2_l$ and $k_{\gamma}^2=0~GeV^2$ for the real photon. Using these simplifications, Eq.[\ref{eq:incomingG}] can be modified as:
	\begin{multline}
	(\mathcal{M}^{\gamma}_{in,SPB})^{\beta}=[\bar{u}(k_2)\dot{\iota}e\gamma^{\beta}u(k_1)]\left(\frac{\dot{\iota}e(k_1\cdot \epsilon^*(k_{\gamma})}{-(k_1\cdot k_{\gamma})}\right),\\
            \\
(\mathcal{M}^{Z}_{in,SPB})^{\beta}=[\bar{u}(k_2)\dot{\iota}e(a_{v}\gamma ^{\beta}+a_p\gamma^{\beta}\gamma_5)u(k_1)]\\
\times\left(\frac{\dot{\iota}e(k_1\cdot \epsilon^*(k_{\gamma})}{-(k_1\cdot k_{\gamma})}\right).
	\label{eq:incomingGfinal}
	\end{multline}
In the similar way, the amplitude for the final state bremsstrahlung $(\mathcal{M}_{fin,SPB})^{\beta}$ can be written as:
    	\begin{multline}
	(\mathcal{M}^{\gamma}_{fin,SPB})^{\beta}=[\bar{u}(k_2)\dot{\iota}e\gamma^{\beta}u(k_1)]\left(\frac{\dot{\iota}e(k_2\cdot \epsilon^*(k_{\gamma})}{(k_2\cdot k_{\gamma})}\right),\\
			\\
           (\mathcal{M}^{Z}_{fin,SPB})^{\beta}=[\bar{u}(k_2)\dot{\iota}e(a_{v}\gamma ^{\beta}+a_p\gamma^{\beta}\gamma_5)u(k_1)]\\
            \times\left(\frac{\dot{\iota}e(k_2\cdot \epsilon^*(k_{\gamma})}{(k_2\cdot k_{\gamma})}\right).
			\label{eq:outgoingGfinal}
	\end{multline}
After taking the amplitude squared and applying polarization sum, we get for $L^{SPB}_{\alpha \beta}$:
	\begin{equation}
            L^{SPB}_{\alpha \beta}=L^{0}_{\alpha \beta}\delta_{SPB},
           \label{eq:total1loopfullEW-brem}
            \end{equation}
where $L^{SPB}_{\alpha \beta}$ is the SPB leptonic tensor and $\delta_{SPB}$ is the SPB factor. The term $L^{0}_{\alpha \beta}$ is the tree-level electroweak leptonic tensor.\\
\indent Integrating on the emitted soft photon phase space, we can write the soft photon factor:
	\begin{equation}
    \begin{split}
			 \delta_{SPB}=-\frac{4\pi \alpha}{2(2\pi)^3}\sum\limits_{i,j=1}^{2}\int_{|\vec{k}_{\gamma}|<\Delta E}\frac{d^3\vec{k}_{\gamma}}{\omega}\frac{(k_i\cdot k_j)}{(k_i\cdot k_{\gamma})(k_j\cdot k_{\gamma})}\\
            \\
			= -\frac{\alpha}{4\pi^2}\sum\limits_{i,j=1}^{2}(k_i\cdot k_j)I(k_i,k_j).
			\label{eq:SPB-phase space integral}
            \end{split}
	\end{equation}
	In Eq.[\ref{eq:SPB-phase space integral}], $\omega$ is the energy of emitted photon, whereas $\Delta E$ is the cut on the soft-photon's energy, which is usually defined by the detector threshold. The soft photon integral $I(k_i,k_j)$ has already been calculated in \cite{THOOFT1979365} and details are given in Appendix \ref{Bremsstrahlung appendix}.\\
\indent Using the example of muon-proton scattering, we demonstrated the cancellation of IR regularization parameter $\lambda$ by adding one-loop corrected and one photon emission $|\mathcal{M}_{SPB}|^2$. The result is shown in Fig.[\ref{fig:musebrem.pdf}].\\
\indent We applied same SPB technique to our NNLO results (quadratic and two-loop reducible contributions). 
 \subsection{Infrared divergence treatment at NNLO}
\begin{figure*}[htb]
		\centering
		\includegraphics[scale=0.5]{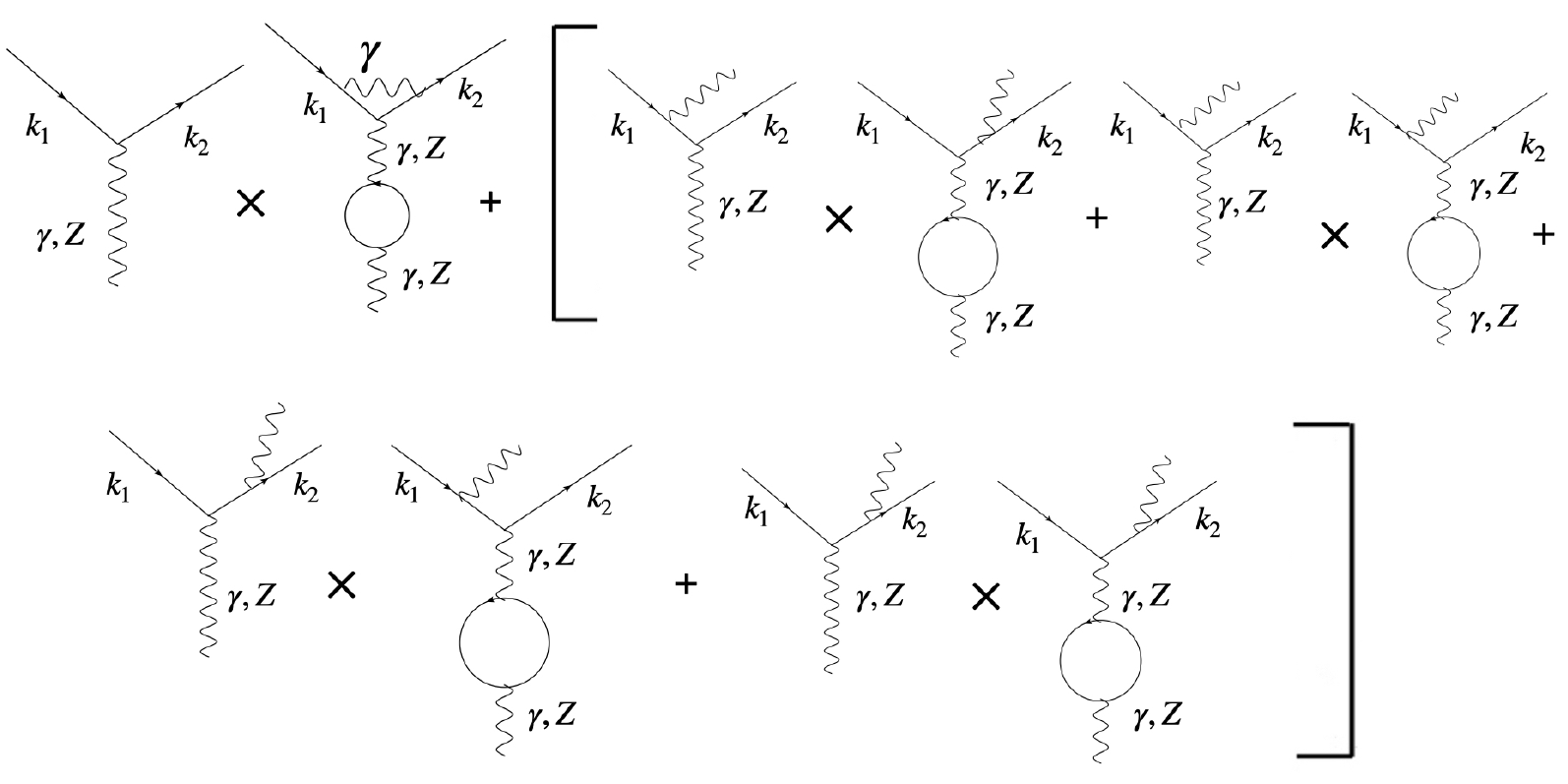}
		\caption{Treatment of IR divergence at reducible two-loop level by including $\Re(\mathcal{M}_{tree}\mathcal{M}^{\dagger}_{SE})$ SPB diagrams.}
		\label{fig:twoloopbremtot.jpeg}
	\end{figure*}

\begin{figure*}[htb]
		\centering
		\includegraphics[scale=0.5]{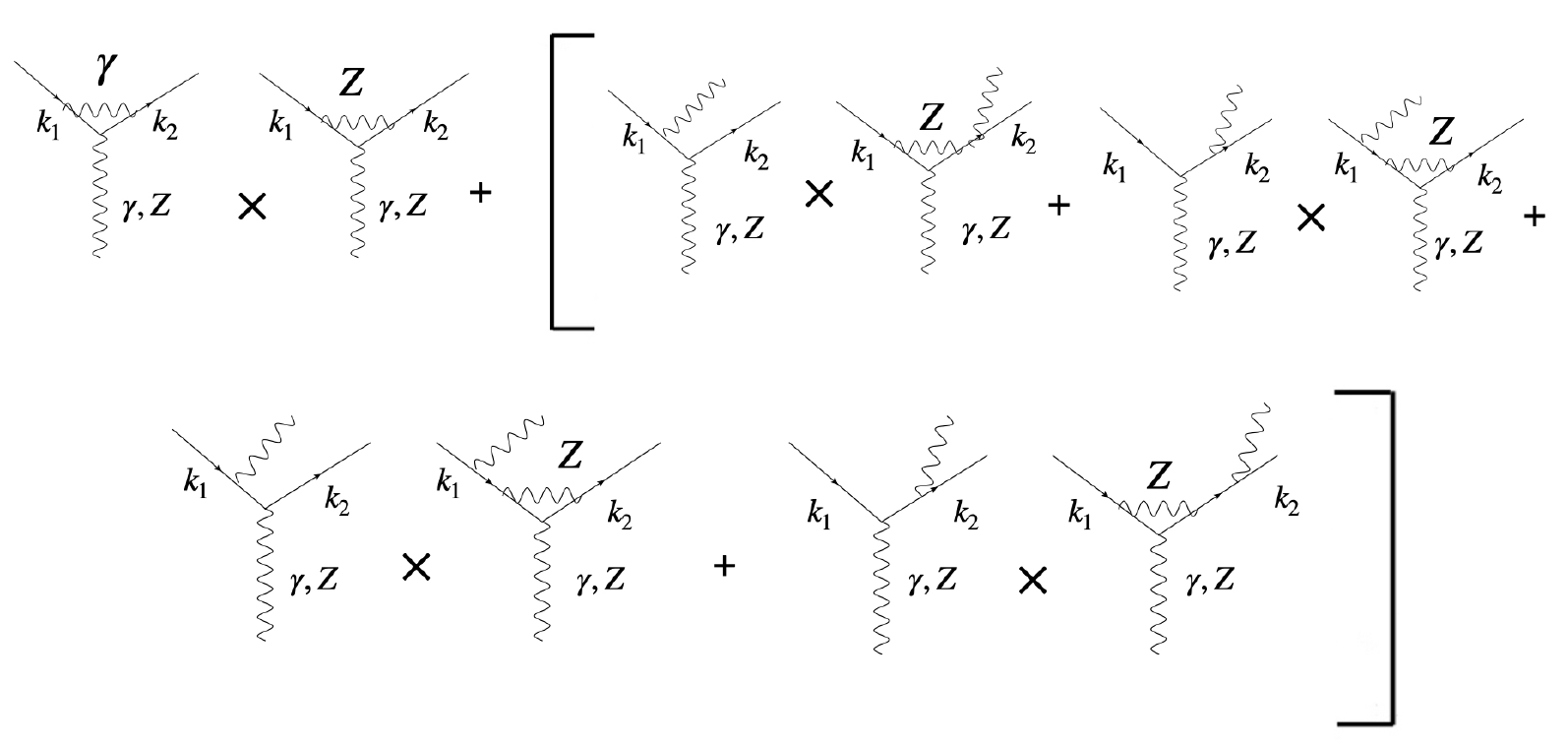}
		\caption{Treatment of IR divergence due to the product of vertex and self-energy graphs at quadratic level by including $\Re(\mathcal{M}_{tree}\times\mathcal{M}^{\dagger}_{Z-vertex})$ SPB diagrams.}
		\label{fig:treeTrbremgraphs.jpeg}
	\end{figure*}

 The IR divergence at the reducible two-loop level can be removed analytically by adding a product of tree-level and one-loop level graphs that both have initial- and final-state soft-photon emission as shown in Fig.[\ref{fig:twoloopbremtot.jpeg}]. 
 In this way, one gets an IR finite two-loop reducible interference term:
	\begin{multline}
		2\Re(\mathcal{M}_{tree}\times\mathcal{M}^{\dagger}_{2lr})_{IR-fin.}=\\
        2\Re(\mathcal{M}_{tree}\times\mathcal{M}^{\dagger}_{2lr})_{IR-div.}+\\
       \left(\frac{1}{2}\right)\times 2\Re (\mathcal{M}_{tree}\times\mathcal{M}^{\dagger}_{SE})\delta_{SP},
		\label{eq:twoloopredIRfinite}
	\end{multline}
	where $\mathcal{M}_{tree}$ and $\mathcal{M}_{SE}$ are the electroweak tree-level and one-loop self-energy amplitudes respectively. Here a factor of $\frac{1}{2}$ appears due to the fact that one half portion of $2\Re
	(\mathcal{M}_{tree}\times \mathcal{M}_{SE})\delta_{SP}$ goes to the IR-divergence treatment of reducible two-loop graphs, whereas the other half goes into the treatment of quadratic graphs.\\
 \indent The NNLO quadratic electroweak lepton scattering involves a product of vertex and self-energy as well as vertex-squared diagrams that contain IR divergence. The IR divergence due to the product of vertex and self-energy graph is removed by adding the product of tree and vertex correction graphs that both have initial- and final-state soft-photon emission as shown in Fig.[\ref{fig:treeTrbremgraphs.jpeg}]. Solving these SPB graphs one gets a one-loop interference term multiplied by a soft-photon factor $\delta_{SP}$, i.e., $2\Re{(\mathcal{M}_{tree}\times\mathcal{M}^{\dagger}_{vertex})}\delta_{SP}$.  
	
\begin{figure*}[!htb]
	\centering
	\includegraphics[scale=0.5]{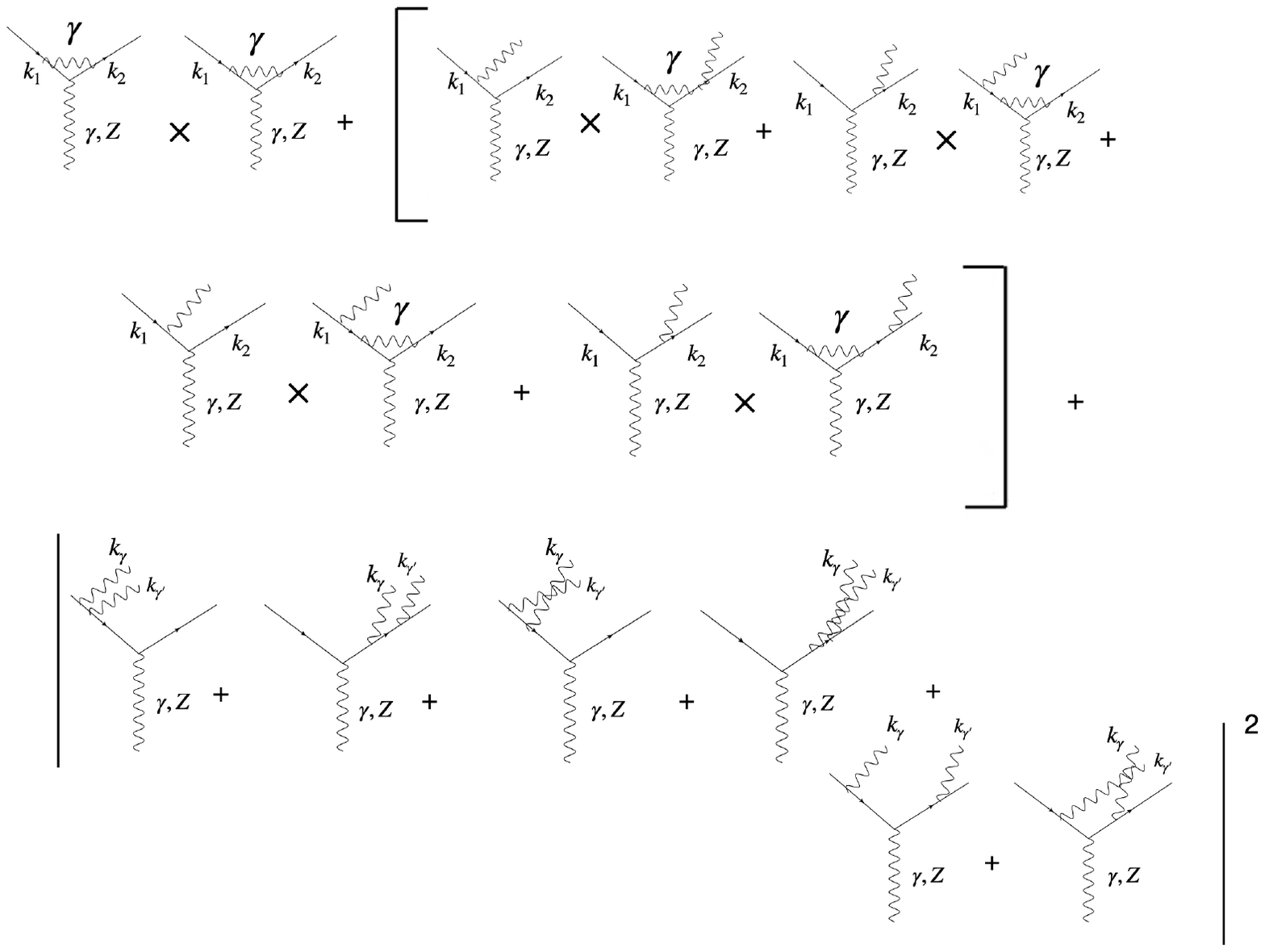}
	\caption{SPB treatment for IR divergence in quadratic NNLO photon vertex-squared contribution. This includes the sum of NLO contribution arising from initial and final state photon emission and two photon emission graphs.}
	\label{fig:treeTrbremgraphs2.jpeg}
\end{figure*}

 However, in order to remove photon vertex-squared IR divergence, one needs to include six two-photon emission diagrams with initial- and final-state radiated photons along with $\Re(\mathcal{M}_{tree}\times\mathcal{M}^{\dagger}_{\gamma-vertex})$ SPB diagrams, as shown in Fig.[\ref{fig:treeTrbremgraphs2.jpeg}].
	
 Hence, the final quadratic level IR finite squared amplitude can be written as:
	\begin{multline}
		|\mathcal{M}_{1l}|^{2}_{IR-fin.}=|\mathcal{M}_{1l}|^{2}_{IR-div.}+\\
\left(\frac{1}{2}\right)\times 2\Re {(\mathcal{M}_{tree}\times\mathcal{M}^{\dagger}_{vertex}+\mathcal{M}_{tree}\times\mathcal{M}^{\dagger}_{SE})}\delta_{SP}+\\
2\Re{(\mathcal{M}_{tree}\times\mathcal{M}^{\dagger}_{vertex})}\delta_{SP}+
		|\mathcal{M}_{tree}|^2\delta^2_{SP},
        \label{eq:quad IR finite amp2}
	\end{multline}
	where $\delta_{SP}^2$ is the soft-photon bremsstrahlung factor squared that we already calculated and $\mathcal{M}_{tree}$, $\mathcal{M}_{SE}$, and $\mathcal{M}_{vertex}$ are the electroweak tree-level, one-loop self-energy, and vertex correction amplitudes, respectively.
 
 \section{numerical analysis}\label{graphical results}
\begin{figure}[!htb]
	\centering
	\includegraphics[scale=0.33]{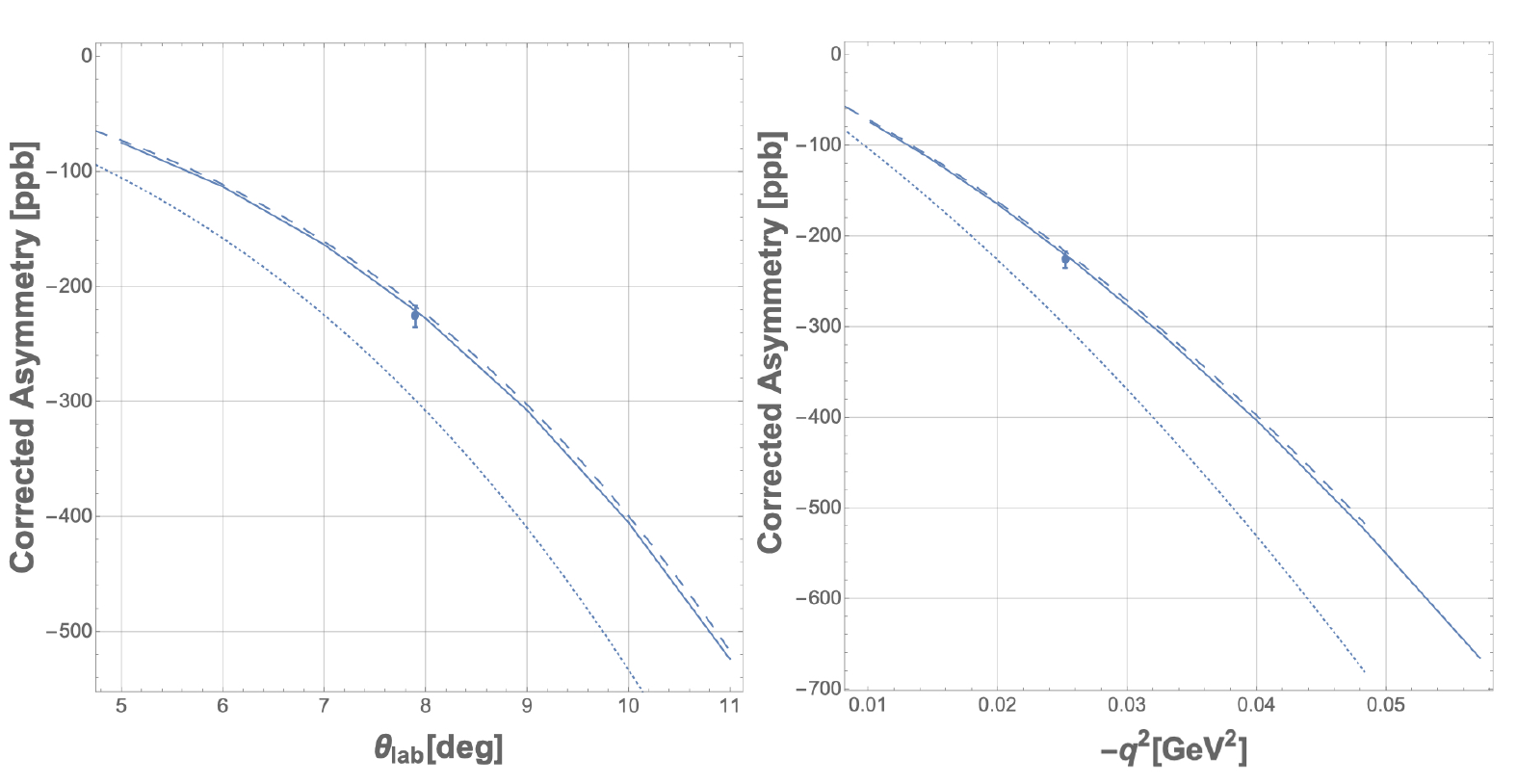}
	\caption{$Q_{weak}$ kinematics: Tree level (dotted line), NLO level (dashed line) and NNLO (quadratic+two-loop reducible) level (solid line) $ep$ scattering correction asymmetry plotted versus $\theta_{lab}$ and $-q^2$. The $Q_{weak}$ measured value at $\theta_{lab}=7.9^0$ is shown by the point with error bars.}
	\label{fig:qweak_asy}
\end{figure}

\begin{figure}[!htb]
	\centering
	\includegraphics[scale=0.33]{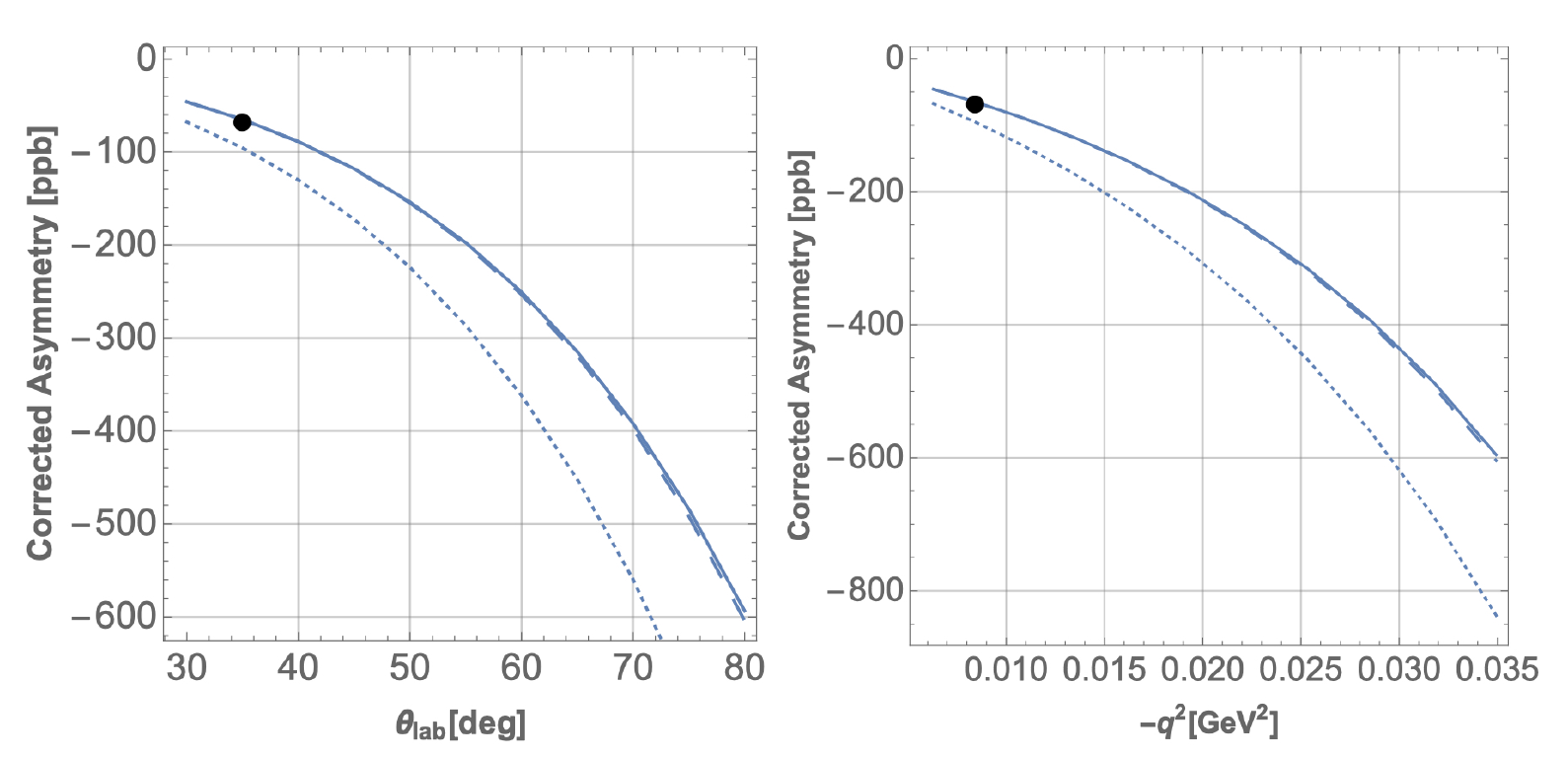}
	\caption{P2 kinematics: Tree level (dotted line), NLO level (dashed line) and NNLO (quadratic+two-loop reducible) level (solid line) $ep$ scattering correction asymmetry plotted versus $\theta_{lab}$ and $-q^2$. The P2 proposed value at $\theta_{lab}=35^0$ is shown by the point.}
	\label{fig:p2_asy}
\end{figure}

\begin{figure}[!htb]
	\centering
	\includegraphics[scale=0.33]{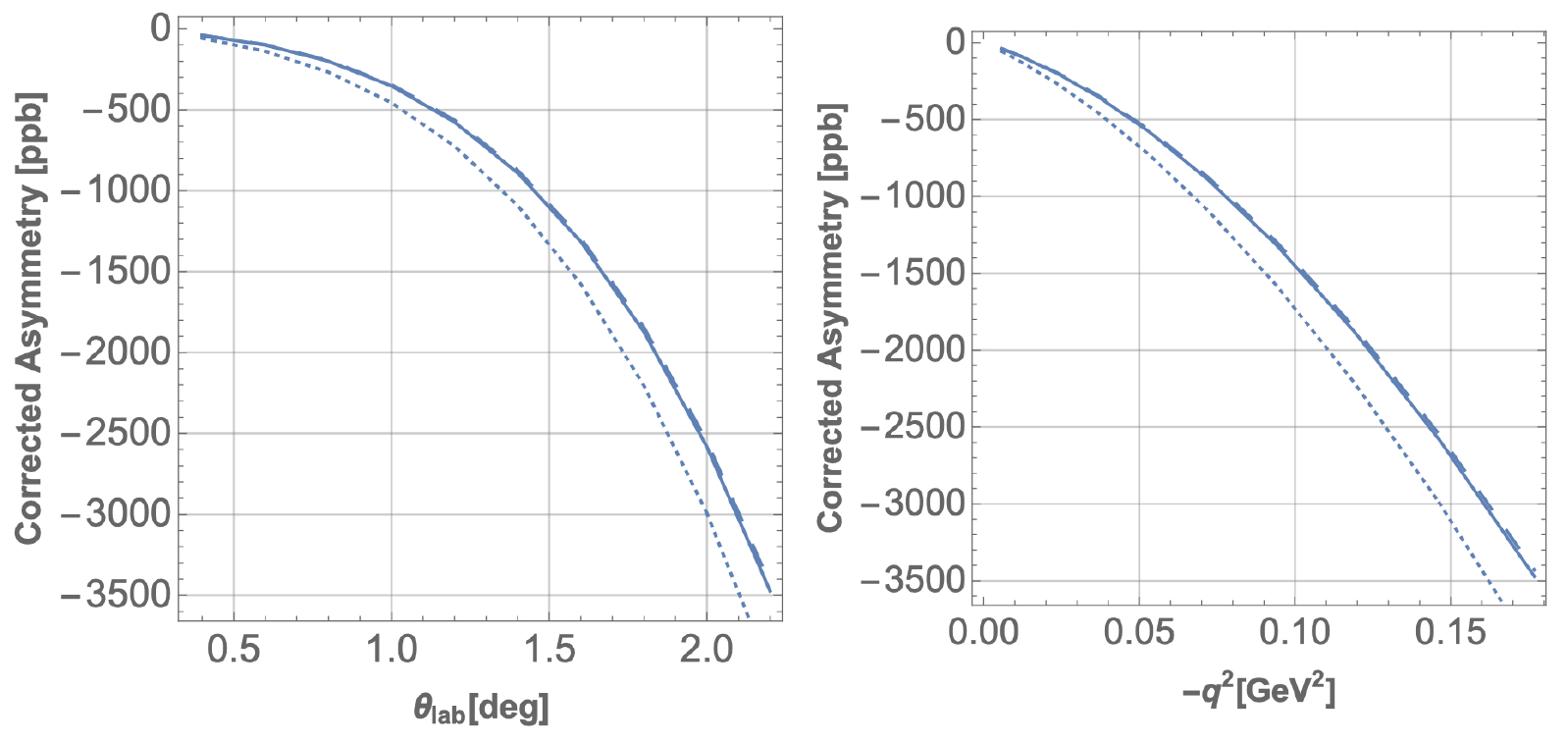}
\caption{MOLLER kinematics: Tree level (dotted line), NLO level (dashed line) and NNLO (quadratic+two-loop reducible) level (solid line) $ep$ scattering correction asymmetry plotted versus $\theta_{lab}$ and $-q^2$.}
	\label{fig:moller_asy}
\end{figure}

\begin{figure}[!htb]
	\centering
	\includegraphics[scale=0.3]{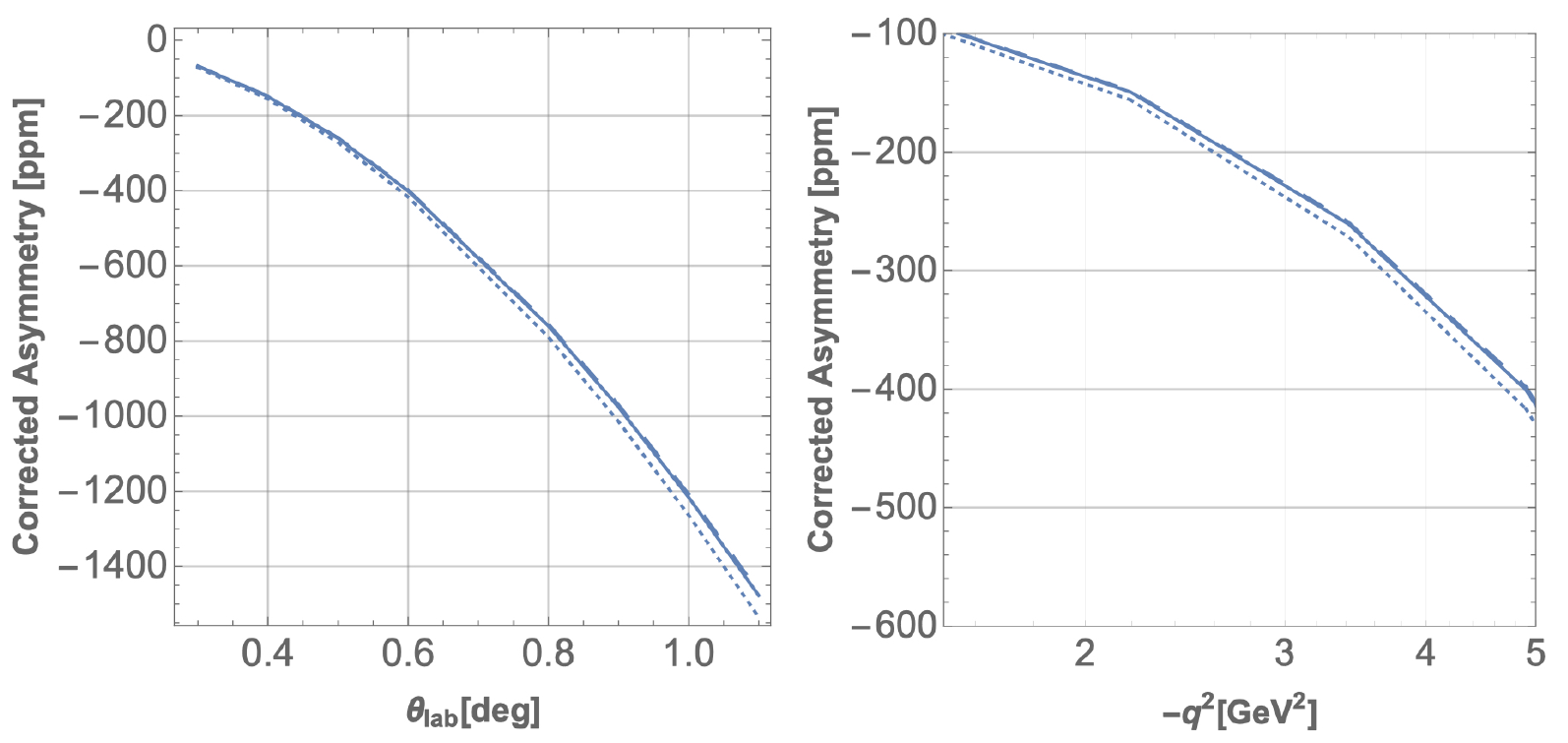}
	\caption{EIC kinematics: Tree level (dotted line), NLO level (dashed line) and NNLO (quadratic+two-loop reducible) level (solid line) $ep$ scattering correction asymmetry plotted versus $\theta_{lab}$ at $E_{CMS}$ = $20~GeV$ and $-q^2$.}
	\label{fig:eic_asy}
\end{figure}

\begin{figure}[!htb]
	\centering
	\includegraphics[scale=0.33]{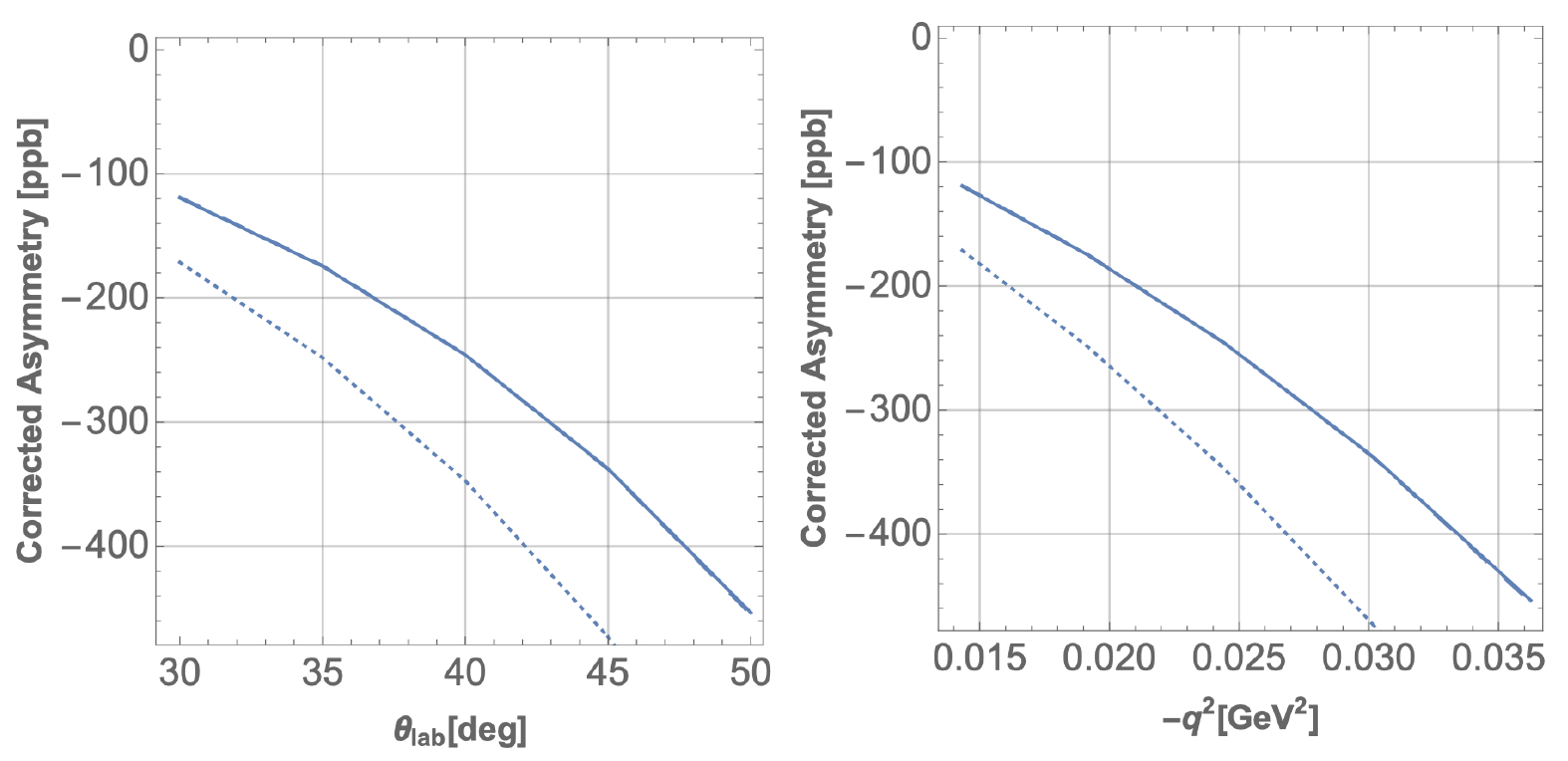}
	\caption{MUSE $(ep)$ scattering  kinematics: Tree level (dotted line), NLO level (dashed line) and NNLO (quadratic+two-loop reducible) level (solid line) $ep$ scattering correction asymmetry plotted versus $\theta_{lab}$ and $-q^2$ at $E_{lab}=0.235~GeV$.}
	\label{fig:Museep_asy}
\end{figure}

\begin{figure}[!htb]
	\centering
	\includegraphics[scale=0.33]{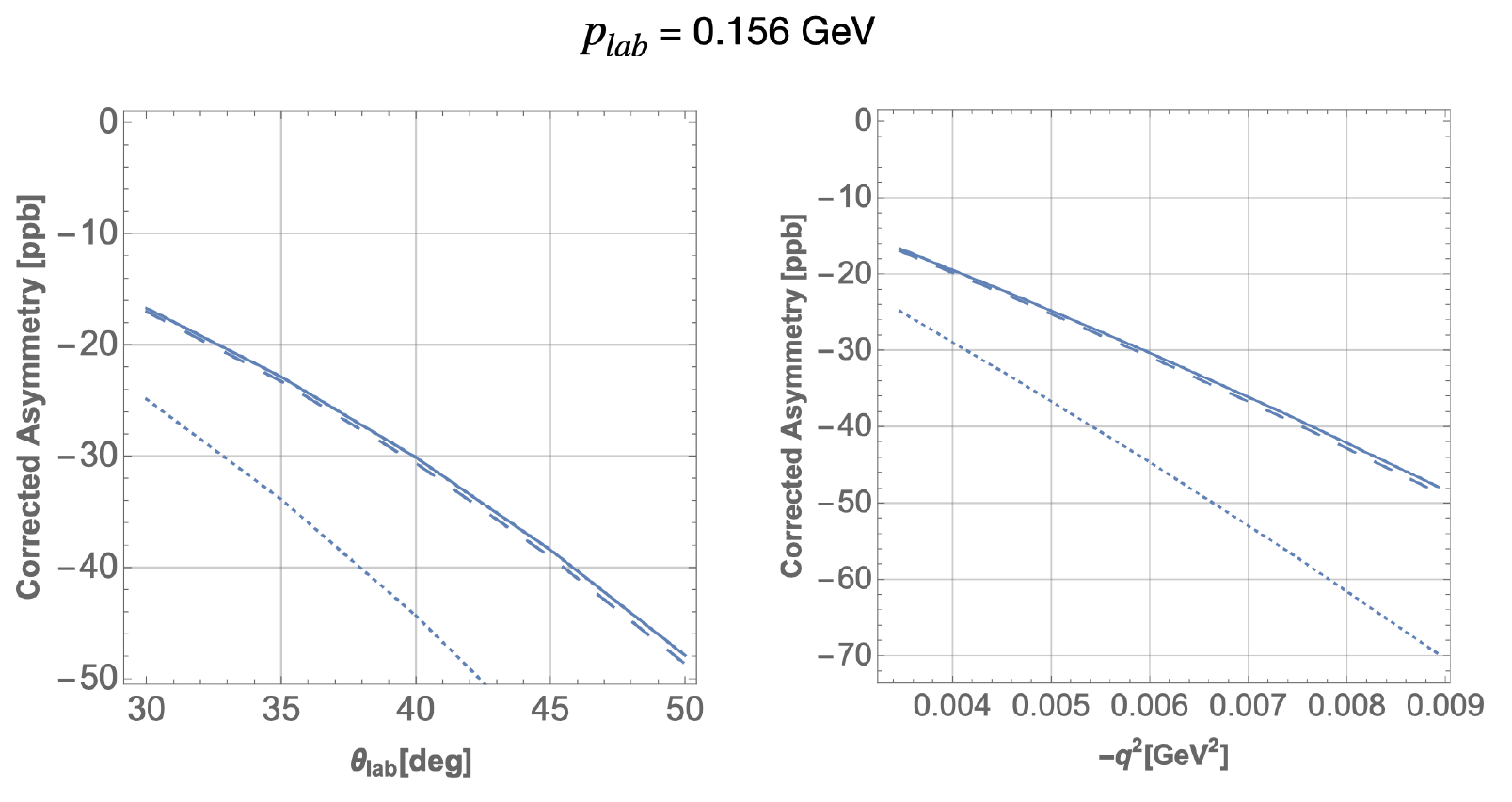}
	\caption{MUSE $(\mu p)$ scattering  kinematics: Tree level (dotted line), NLO level (dashed line) and NNLO (quadratic+two-loop reducible) level (solid line) $\mu p$ scattering correction asymmetry plotted versus $\theta_{lab}$ and $-q^2$ at $E_{lab}$ = $0.156~GeV$.}
	\label{fig:Muse_asy1}
\end{figure}

\begin{figure}[!htb]
	\centering
	\includegraphics[scale=0.33]{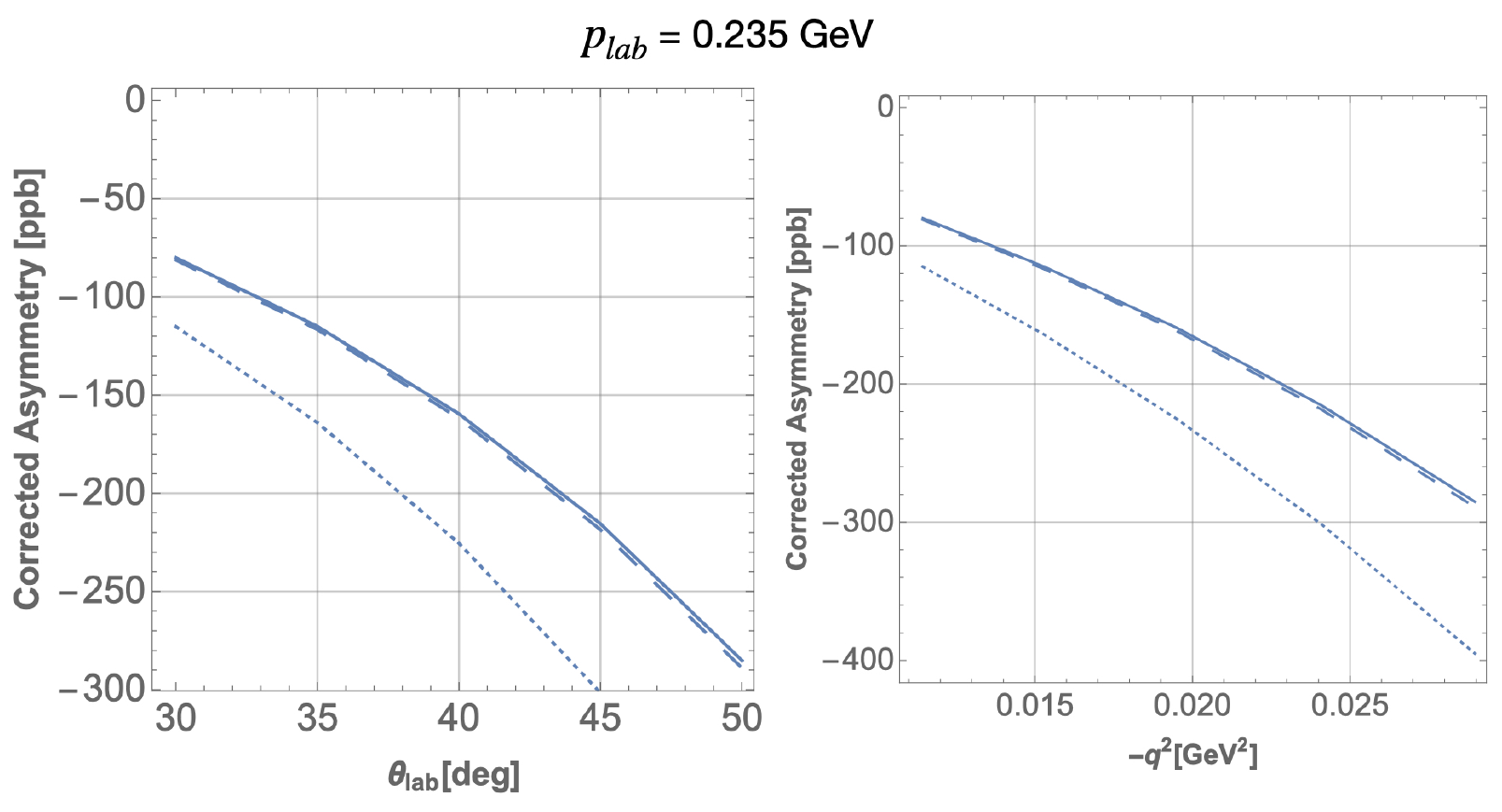}
	\caption{MUSE $(\mu p)$ scattering  kinematics: Tree level (dotted line), NLO level (dashed line) and NNLO (quadratic+two-loop reducible) level (solid line) $\mu p$ scattering correction asymmetry plotted versus $\theta_{lab}$ and $-q^2$ at $E_{lab}$ = $0.235~GeV$.}
	\label{fig:Muse_asy2}
\end{figure}

We have calculated the tree-level, NLO, and NNLO (tree + one-loop + quadratic + reducible two-loop) electroweak parity-violating asymmetry $(A_{PV})$ for the lepton-proton elastic scattering by using the covariant approach. The Feynman diagrams are generated using FeynArts mathematica package. The leptonic and hadronic tensors are calculated using FormCalc, which are then contracted to obtain the squared amplitudes with FeynCalc package.\\ 
\indent We compared our NLO and NNLO $(A_{PV})$ results with the measured $Q_{weak}$ experimental value. Our results are in good agreement, but we need to account boxes and hard photon bremsstrahlung cross-section. The NLO and NNLO corrected amplitude squared $(|\mathcal{M}|^2)$ is given below:
\begin{multline}
		|\mathcal{M}^{0+1}_{NLO}|^2=\mathcal|{M}_{tree}|^2+2\Re (\mathcal{M}_{tree}\times \mathcal{M}^\dagger_{1l}),\\
        \\
		|\mathcal{M}^{0+1+2}_{NNLO}|^2=|\mathcal{M}^{0+1}_{NLO}|^2+|\mathcal{M}_{1l}|^2+2\Re(\mathcal{M}_{tree}\times\mathcal{M}^{\dagger}_{2lr}).
			\label{eq:qud and two loop red definitions}
	\end{multline}
where in Eq.[\ref{eq:qud and two loop red definitions}], the superscripts $0,~1$ and $2$ are used for up to the LO, NLO and NNLO contributions in amplitude squared. The term $|\mathcal{M}_{1l}|^2$ represents the one-loop level squared amplitude which is the quadratic contribution. The term $2\Re(\mathcal{M}_{tree}\times\mathcal{M}^{\dagger}_{2lr})$
is for the reducible two-loop interference term.\\
\indent Our numerical results for elastic $lp$ scattering using the kinematics of the experimental programs of $Q_{weak}$, P2, MOLLER, EIC and MUSE are shown in Figs.[\ref{fig:qweak_asy}-\ref{fig:Muse_asy2}]. These graphs show the tree-level $A_{PV}$ along with the corrections in $A_{PV}$ at the NLO and NNLO levels. The soft photon bremsstrahlung (SPB) results with the emission of one and two photons are also added in these graphs to make them IR finite. These $A_{PV}$ graphs are plotted versus both the momentum transfer $-q^2$ and the scattering angle $(\theta_{lab})$ in the lab reference frame. In our calculations, we take the soft photon energy cut as $\Delta E= 0.05\sqrt{s}~GeV$ with $s$ being the center-of-mass energy squared i.e., $s=E_{CMS}^2$.  


\subsection{$Q_{weak}$ kinematics}\label{q-weak results}
In case of the $Q_{weak}$ experiment, Fig.[\ref{fig:qweak_asy}] shows our numerical results with the beam energy $E_{beam}=1.16~GeV$. At this $E_{beam}$, the $Q_{weak}$ measured value of $A_{PV}$ at $\theta_{lab}=7.9^0$ is $-226.5\pm 7.3 (statistical)\pm 5.8(systematic)~parts~per~billion~(ppb)$ \cite{Qweak:2013zxf}. Using these kinematics, our calculated values of the corrected asymmetries at the tree, NLO, and NNLO are given in Table[\ref{table:1}]. We also calculate the $A_{PV}$ correction percentage at NLO $(\delta^1_{A_{PV}})$ and NNLO $(\delta^2_{A_{PV}})$ (quadratic and two-loop reducible) as given by the formulas:
\begin{multline}
			up~to~NLO~correction~(\delta^1_{A_{PV}}\%)=\left(\frac{A_{PV}^{0}-A_{PV}^{0+1}}{A_{PV}^{0}}\right)\times 100,\\
		up~to~NNLO~correction~(\delta^2_{A_{PV}}\%)=\left(\frac{A_{PV}^{0}-A_{PV}^{0+1+2}}{A_{PV}^{0}}\right)\\
        \times 100.
        \label{eq:delta correction per}
\end{multline}

The NLO and NNLO $A_{PV}$ correction percentages $\delta^1_{A_{PV}}\%$ and $\delta^2_{A_{PV}}\%$ for $Q_{weak}$ kinematics at different $\theta_{lab}$ are given in Table[\ref{table:qweak correction per}]. These results are calculated in parts per billion $(ppb)$.

\begin{table}[!h]
	\centering
     \resizebox{\columnwidth}{!}{%
	\begin{tabular}{||c| c| c| c| c|c ||} 
		\hline
		$\theta_{lab}$ & Tree~$A_{PV}(ppb)$  & 1-loop~$A_{PV}(ppb)$ & Qud-$A_{PV}(ppb)$ & 2-loop~red~$A_{PV}(ppb)$ & Total~$A_{PV}(ppb)$ \\ [0.1ex] 
		\hline\hline
		$5^0$ & -105.61 & -73.06 & -74.47 & -73.61 & -75.02 \\ 
		$6^0$ & -158.15 & -111.34 & -112.48 & -112.12 & -113.27  \\
		$7^0$ & -224.89 & -160.99 & -162.61 & -162.04 & -163.66 \\
		$7.9^0$ & -298.90 & -217.11 & -219.89 & -218.40 & -221.19\\
		$8^0$ & -308.02 & -224.08 & -226.49 & -225.41 & -227.83 \\
		$9^0$ & -409.93 & -302.82 & -306.06 & -304.43 & -307.68 \\
		$10^0$ & -533.22 & -399.59 & -403.31 & -401.49 & -405.21 \\
		$11^0$ & -680.63 & -516.94 & -521.81 & -519.09 & -523.97 \\
		$12^0$ & -854.99 & -657.45 & -663.34 & -659.82 & -665.73 \\
		$13^0$ & -1059.18 & -823.79 & -830.75 & -826.33 & -833.31 \\
		$14^0$ & -1296.08 & -1018.60 & -1026.69 & -1021.25 & -1029.36 \\
		$15^0$ & -1568.53 & -1244.52 & -1253.76 & -1247.17 & -1256.44 \\[0.1ex] 
		\hline
	\end{tabular}%
    }
	\caption{$Q_{weak}$ kinematics: Our calculated tree, one-loop, quadratic, reducible two-loop and total (quadratic+reducible two-loop) corrected $A_{PV}$ at different scattering angles $\theta_{lab}$. Here (Qud.) represents the quadratic and (2-loop~red.) is for reducible two-loop corrected $A_{PV}$. These results are calculated in $ppb$.}
	\label{table:1}
\end{table}

\begin{figure}[!htb]
	\centering
	\includegraphics[scale=0.5]{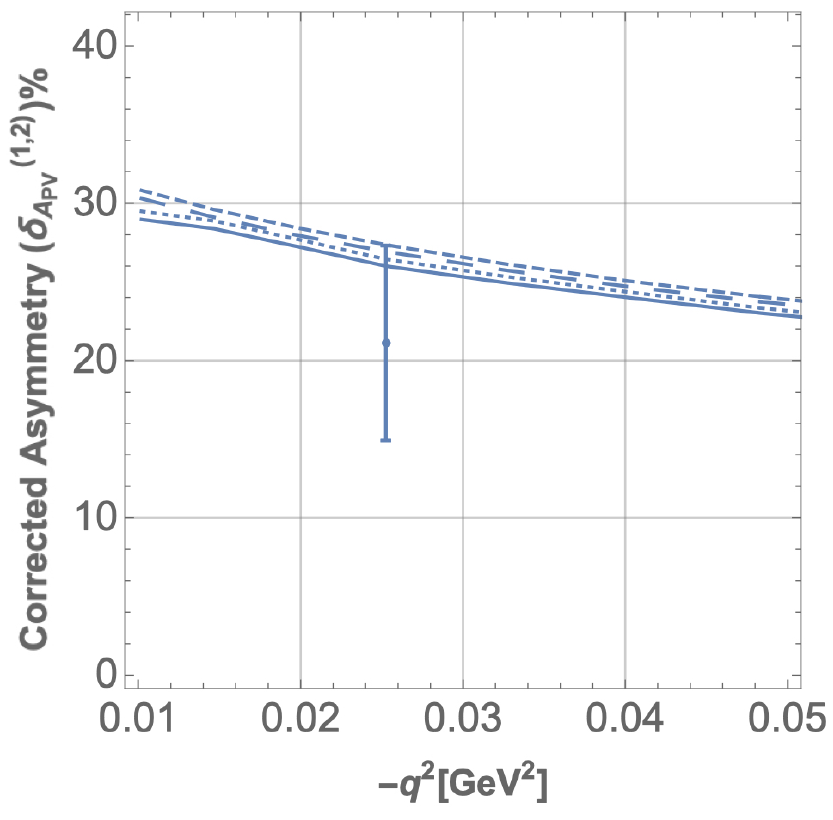}
	\caption{$Q_{weak}$ kinematics: Tree+NLO $\delta^1_{A_{PV}}\%$ (dashed line), Tree+NLO+quadratic $\delta^2_{A_{PV}}\%$ (dotted line), Tree+NLO+two-loop reducible $\delta^2_{A_{PV}}\%$ (large-dashed line) and Tree+NLO+quadratic+two-loop reducible $\delta^2_{A_{PV}}\%$ (solid line) plotted versus $-q^2$. The point with error bar represents the $Q_{weak}$ measured corrected $A_{PV}$ at $-q^2=0.025~GeV^2.$ Here $\delta^1_{A_{PV}}\%$ and $\delta^2_{A_{PV}}\%$ are NLO and NNLO $A_{PV}$ correction percentages as given by Eq.[\ref{eq:delta correction per}].}
	\label{fig:qweak correction percent}
\end{figure}

\begin{table}[!h]
	\centering
     \resizebox{\columnwidth}{!}{%
	\begin{tabular}{||c| c| c| c| c ||} 
		\hline
		$\theta_{lab}$ & $\delta^1_{A_{PV}}(1~loop)\%$  & $\delta^2_{A_{PV}}(qud)\%$ & $\delta^2_{A_{PV}}(2~loop~red)\%$ & $\delta^2_{A_{PV}}(qud+2~loop~red)\%$ \\ [0.5ex] 
		\hline\hline
		$5^0$ & 30.82 & 29.49& 30.30 & 28.97 \\ 
		$6^0$ &29.60& 28.88 & 29.11 & 28.38 \\
		$7^0$ & 28.41 & 27.69 &27.95 & 27.23 \\
		$7.9^0$ & 27.37 & 26.44 & 26.93 & 26.00 \\
		$8^0$ &27.25 & 26.47 & 26.82 & 26.03 \\
		$9^0$ & 26.13 &25.34 & 25.73 & 24.94 \\
		$10^0$ & 25.06& 24.36 & 24.70 & 24.01 \\
		$11^0$ & 24.05 & 23.33 & 23.73 & 23.02 \\
		$12^0$ & 23.10 & 22.42 & 22.83 & 22.14 \\
		$13^0$ & 22.22 & 21.57 &21.98 & 21.32 \\
		$14^0$ & 21.41 & 20.78 & 21.21 & 20.58 \\
		$15^0$ & 20.66 & 20.07 & 20.49 & 19.89 \\[0.5ex]
		\hline
	\end{tabular}%
    }
	\caption{$Q_{weak}$ kinematics: $\delta^1_{A_{PV}}\%$ and $\delta^2_{A_{PV}}\%$ at different scattering angles $\theta_{lab}$. Here $\delta^1_{A_{PV}}\%$ and $\delta^2_{A_{PV}}\%$ are NLO and NNLO $A_{PV}$ correction percentages as given by Eq.[\ref{eq:delta correction per}], whereas (qud.) represents the quadratic and (2 loop~red.) is for reducible two-loop level correction$\%$.}
	\label{table:qweak correction per}
\end{table}
The graphs for $\delta^1_{A_{PV}}\%$ and $\delta^2_{A_{PV}}\%$ plotted versus different $-q^2$ values in the case of $Q_{weak}$ kinematics are shown in Fig.[\ref{fig:qweak correction percent}].

\subsection{P2 kinematics}\label{p2 results}
Using the experimental kinematics of the proposed P2 experiment, our corrected $A_{PV}$ results at tree, NLO, and NNLO plotted versus $-q^2$ and $\theta_{lab}$ are shown in Fig.[\ref{fig:p2_asy}]. This experiment will operate at a very low beam energy $E_{beam}=155~MeV$, making it possible to measure the scattered particles at large scattering angles. The value of $A_{PV}$ proposed by the recent P2 collaborations at $\theta_{lab}=35^0$ is $-67.34~ppb$ \cite{P2.exp}. Our calculated values of $A_{PV}$ at tree, NLO, and NNLO for this angle are given in Table[\ref{table:2}].

\begin{table}[!h]
	\centering
     \resizebox{\columnwidth}{!}{%
	\begin{tabular}{||c| c| c| c| c|c ||} 
		\hline
		$\theta_{lab}$ & Tree~$A_{PV}(ppb)$  & 1-loop~$A_{PV}(ppb)$ & Qud-$A_{PV}(ppb)$ & 2-loop~red~$A_{PV}(ppb)$ & Total~$A_{PV}(ppb)$ \\ [0.5ex] 
		\hline\hline
		$30^0$ & -67.80 & -45.98 & -46.31 & -45.70 & -46.03 \\ 
		$35^0$ & -95.61 & -65.14 & -65.56 & -64.67 & -65.09  \\
		$40^0$ & -130.08 & -89.04 & -89.57 & -88.30 & -88.83 \\
		$45^0$ & -172.40 & -118.59 & -119.23 & -117.46 & -118.10\\
		$50^0$ & -223.99 & -154.91 & -155.66 & -153.25 & -154.00 \\
		$55^0$ & -286.52 & -199.33 & -200.19 & -196.97 & -197.82 \\
		$60^0$ & -361.88 & -253.39 & -254.37 & -250.12 & -251.08 \\
		$65^0$ & -452.19 & -318.89 & -319.95 & -314.45 & -315.49 \\
		$70^0$ & -559.81 & -397.75 & -398.88 & -391.86 & -392.97 \\
		$75^0$ & -687.19 & -492.07 & -493.26 & -484.42 & -485.58 \\
		$80^0$ & -836.91 & -604.07 & -605.29 & -594.31 & -595.48 \\[1ex] 
		\hline
	\end{tabular}%
    }
	\caption{P2 kinematics: Our calculated tree, one-loop, quadratic, reducible two-loop and total (quadratic+reducible two-loop) corrected $A_{PV}$ at different scattering angles $\theta_{lab}$. Here (Qud.) represents the quadratic and (2-loop~red.) is for reducible two-loop level corrected $A_{PV}$. These results are calculated in $ppb$.}
	\label{table:2}
\end{table}

\begin{figure}[htb]
	\centering
	\includegraphics[scale=0.45]{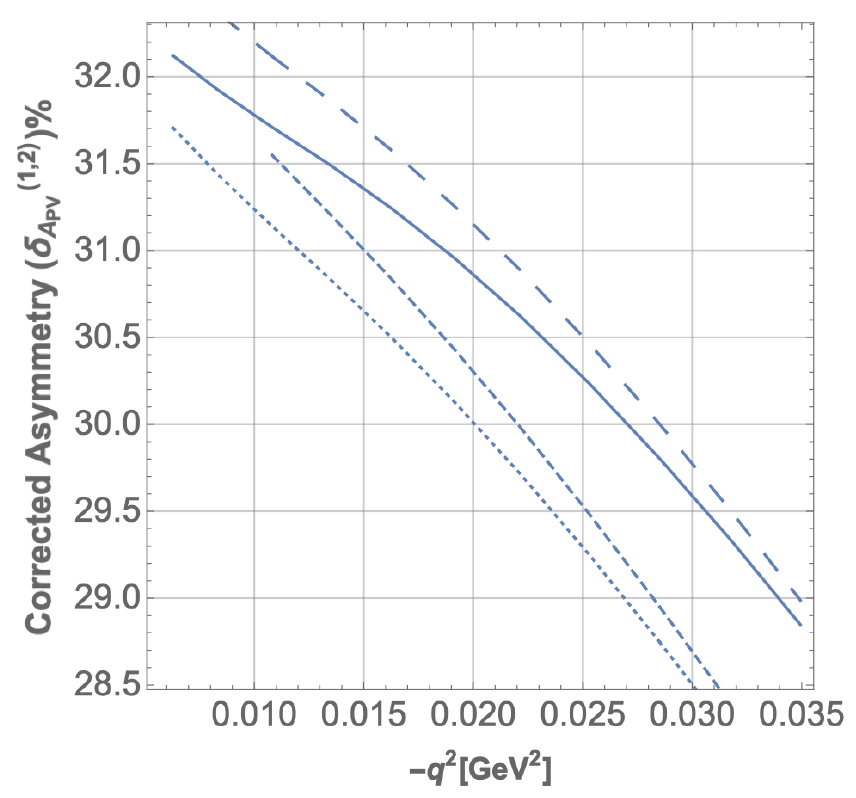}
	\caption{P2 kinematics: Tree+NLO $\delta^1_{A_{PV}}\%$ (dashed line), Tree+NLO+quadratic $\delta^2_{A_{PV}}\%$ (dotted line), Tree+NLO+two-loop reducible $\delta^2_{A_{PV}}\%$ (large-dashed line) and Tree+NLO+quadratic+two-loop reducible $\delta^2_{A_{PV}}\%$ (solid line) plotted versus $-q^2$. Here $\delta^1_{A_{PV}}\%$ and $\delta^2_{A_{PV}}\%$ are NLO and NNLO $A_{PV}$ correction percentages as given by Eq.[\ref{eq:delta correction per}].}
	\label{fig:p2 correction percent}
\end{figure}

The NLO and NNLO $A_{PV}$ correction percentages $\delta^1_{A_{PV}}\%$ and $\delta^2_{A_{PV}}\%$ for P2 kinematics are given in Table[\ref{table:p2 correction per}] and shown graphically in Fig.[\ref{fig:p2 correction percent}].

\begin{table}[!h]
	\centering
     \resizebox{\columnwidth}{!}{%
	\begin{tabular}{||c| c| c| c| c ||} 
		\hline
		$\theta_{lab}$ & $\delta^1_{A_{PV}}(1~loop)\%$  & $\delta^2_{A_{PV}}(qud)\%$ & $\delta^2_{A_{PV}}(2~loop~red)\%$ & $\delta^2_{A_{PV}}(qud+2~loop~red)\%$ \\ [0.5ex] 
		\hline\hline
		$30^0$ & 32.18 & 31.70& 32.59 & 32.12\\ 
		$35^0$ &31.87 & 31.43 & 32.36 & 31.91\\
		$40^0$ & 31.55 & 31.14 & 32.12 & 31.71 \\
		$45^0$ & 31.21 & 30.84 & 31.87 & 31.49 \\
		$50^0$ & 30.84 & 30.51 & 31.58 & 31.25 \\
		$55^0$ & 30.43 & 30.13 & 31.26 & 30.96\\
		$60^0$ & 29.98 & 29.71 & 30.88 & 30.62 \\
		$65^0$ & 29.48 & 29.24 & 30.46 & 30.23 \\
		$70^0$ & 28.95 & 28.75 & 30.00 & 29.80 \\
		$75^0$ & 28.39 & 28.22 & 29.51 & 29.34 \\
		$80^0$ & 27.82 & 27.68 & 28.99 & 28.85 \\[1ex]
		\hline
	\end{tabular}%
    }
	\caption{P2 kinematics: $\delta^1_{A_{PV}}\%$ and $\delta^2_{A_{PV}}\%$ at different scattering angles $\theta_{lab}$. Here $\delta^1_{A_{PV}}\%$ and $\delta^2_{A_{PV}}\%$ are NLO and NNLO $A_{PV}$ correction percentages as given by Eq.[\ref{eq:delta correction per}], whereas (qud.) represents the quadratic and (2 loop~red.) is for reducible two-loop level correction$\%$.}
	\label{table:p2 correction per}
\end{table}

\subsection{MOLLER kinematics}\label{moller results}
The MOLLER experiment will measure the parity-violating asymmetry in the scattering of longitudinally polarized electrons off unpolarized electrons in a liquid hydrogen target. In this scenario $ep$ scattering is a significant background process for which one needs to account. \\
\indent Our numerical results for this background ($ep$) scattering process using the MOLLER experimental kinematics at $E_{lab}=11~GeV$ are plotted in Fig.[\ref{fig:moller_asy}] versus different scattering angles $\theta_{lab}$ and momentum transfer squared $(-q^2)$. The $A_{PV}$ values at tree, one-loop, quadratic, two-loop reducible, and total (quadratic + two-loop reducible) with respect to $\theta_{lab}\sim (0.4^0-2.2^0)$ proposed by the MOLLER collaboration \cite{MOLLER:2014iki} are given in Table[\ref{table:moller correction}]. These values are calculated in parts per billion $(ppb)$.
\begin{table}[!h]
	\centering
     \resizebox{\columnwidth}{!}{%
	\begin{tabular}{||c| c| c| c| c|c ||} 
		\hline
		$\theta_{lab}$ & Tree~$A_{PV}(ppb)$  & 1-loop~$A_{PV}(ppb)$ & Qud-$A_{PV}(ppb)$ & 2-loop~red~$A_{PV}(ppb)$ & Total~$A_{PV}(ppb)$ \\ [0.5ex] 
		\hline\hline
			$0.4^0$ & -58.10 & -39.12 & -39.85 & -39.61 & -40.34 \\ 
		$0.6^0$ & -138.92 & -97.26 & -99.08 & -98.42 & -100.24  \\
		$0.8^0$ & -267.03 & -194.19 & -197.38 & -196.32 & -199.51 \\
		$1.0^0$ & -456.59 & -344.12 & -349.03 & -347.55 & -352.46\\
		$1.2^0$ & -724.82 & -564.13 & -571.08 & -569.20 & -576.16\\
		$1.4^0$ & -1091.22 & -873.53 & -882.84 & -880.64 & -889.97\\
		$1.6^0$ & -1576.93 & -1293.15 & -1305.16 & -1302.72 & -1314.76 \\
		$1.8^0$ & -2203.99 & -1844.67 & -1859.70 & -1857.16 & -1872.23 \\
		$2.0^0$ & -2994.61 & -2549.88 & -2568.26 & -2565.79 & -2584.25 \\
		$2.2^0$ & -3970.56 & -3430.12 & -3452.20 & -3450.04 & -3472.22 \\[1ex] 
		\hline
	\end{tabular}%
    }
	\caption{MOLLER kinematics: Our calculated tree, one-loop, quadratic, reducible two-loop and total (quadratic+reducible two-loop) corrected $A_{PV}$ at different scattering angles $\theta_{lab}$. Here (Qud.) represents the quadratic and (2-loop~red.) is for reducible two-loop level corrected $A_{PV}$. These results are calculated in $ppb$.}
	\label{table:moller correction}
\end{table}

\begin{figure}[!htb]
	\centering
	\includegraphics[scale=0.5]{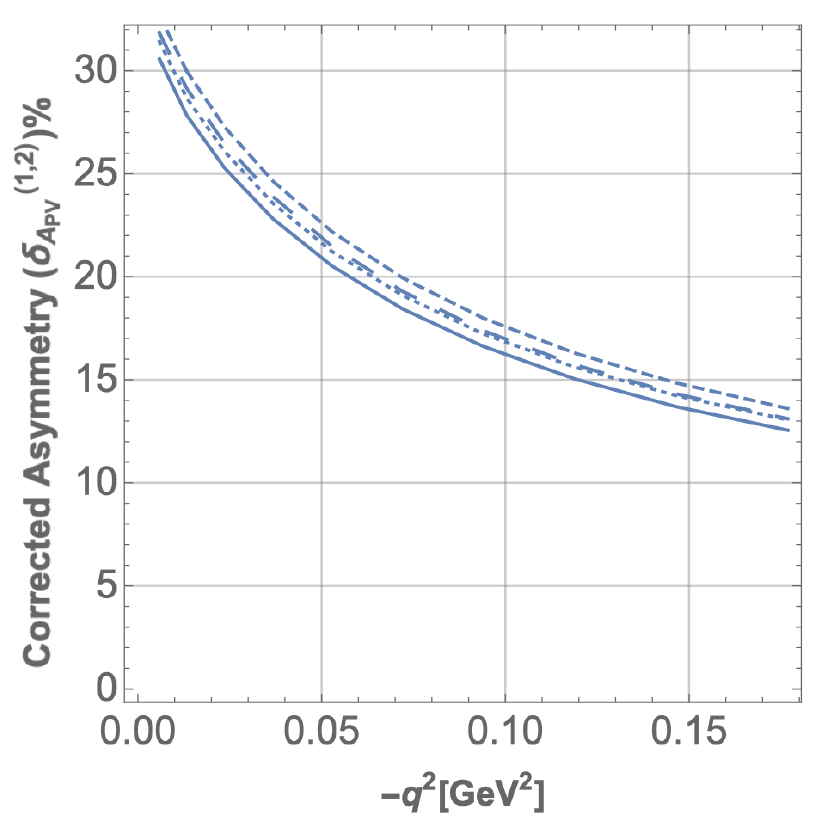}
	\caption{MOLLER kinematics: Tree+NLO $\delta^1_{A_{PV}}\%$ (dashed line), Tree+NLO+quadratic $\delta^2_{A_{PV}}\%$ (dotted line), Tree+NLO+two-loop reducible $\delta^2_{A_{PV}}\%$ (large-dashed line) and Tree+NLO+quadratic+two-loop reducible $\delta^2_{A_{PV}}\%$ (solid line) plotted versus $-q^2$. Here $\delta^1_{A_{PV}}\%$ and $\delta^2_{A_{PV}}\%$ are NLO and NNLO $A_{PV}$ correction percentages as given by Eq.[\ref{eq:delta correction per}].}
	\label{fig:moller correction percent}
\end{figure}

The NLO and NNLO $A_{PV}$ correction percentages $\delta^1_{A_{PV}}\%$ and $\delta^2_{A_{PV}}\%$ for MOLLER kinematics at different scattering angles $(\theta_{lab})$ are given in Table[\ref{table:moller correction per}] and shown graphically in Fig.[\ref{fig:moller correction percent}].

\begin{table}[!h]
	\centering
     \resizebox{\columnwidth}{!}{%
	\begin{tabular}{||c| c| c| c| c ||} 
		\hline
		$\theta_{lab}$ & $\delta^1_{A_{PV}}(1~loop)\%$  & $\delta^2_{A_{PV}}(qud)\%$ & $\delta^2_{A_{PV}}(2~loop~red)\%$ & $\delta^2_{A_{PV}}(qud+2~loop~red)\%$ \\ [0.5ex] 
		\hline\hline
		$0.4^0$ & 32.67 & 31.41 & 31.82 & 30.56 \\ 
		$0.6^0$ & 29.99 & 28.68 & 29.15 & 27.85 \\
		$0.8^0$ & 27.28 & 26.08 & 26.48 & 25.28 \\
		$1.0^0$ & 24.63 & 23.56 & 23.88 & 22.81 \\
		$1.2^0$ & 22.17 & 21.21 & 21.47 & 20.51\\
		$1.4^0$ & 19.95 & 19.09 & 19.29 & 18.44\\
		$1.6^0$ & 17.99 & 17.23 & 17.39 & 16.63  \\
		$1.8^0$ & 16.30 & 15.62 & 15.74 & 15.05  \\
		$2.0^0$ & 14.85 & 14.24 & 14.32 & 13.70 \\
		$2.2^0$ & 13.61 & 13.06 & 13.11 & 12.55  \\[1ex] 
		\hline
	\end{tabular}%
    }
	\caption{MOLLER kinematics: $\delta^1_{A_{PV}}\%$ and $\delta^2_{A_{PV}}\%$ at different scattering angles $\theta_{lab}$. Here $\delta^1_{A_{PV}}\%$ and $\delta^2_{A_{PV}}\%$ are NLO and NNLO $A_{PV}$ correction percentages as given by Eq.[\ref{eq:delta correction per}], whereas (qud.) represents the quadratic and (2 loop~red.) is for reducible two-loop level correction$\%$.}
	\label{table:moller correction per}
\end{table}

\subsection{EIC kinematics}\label{eic results}
\begin{figure}[!htb]
	\centering
	\includegraphics[scale=0.5]{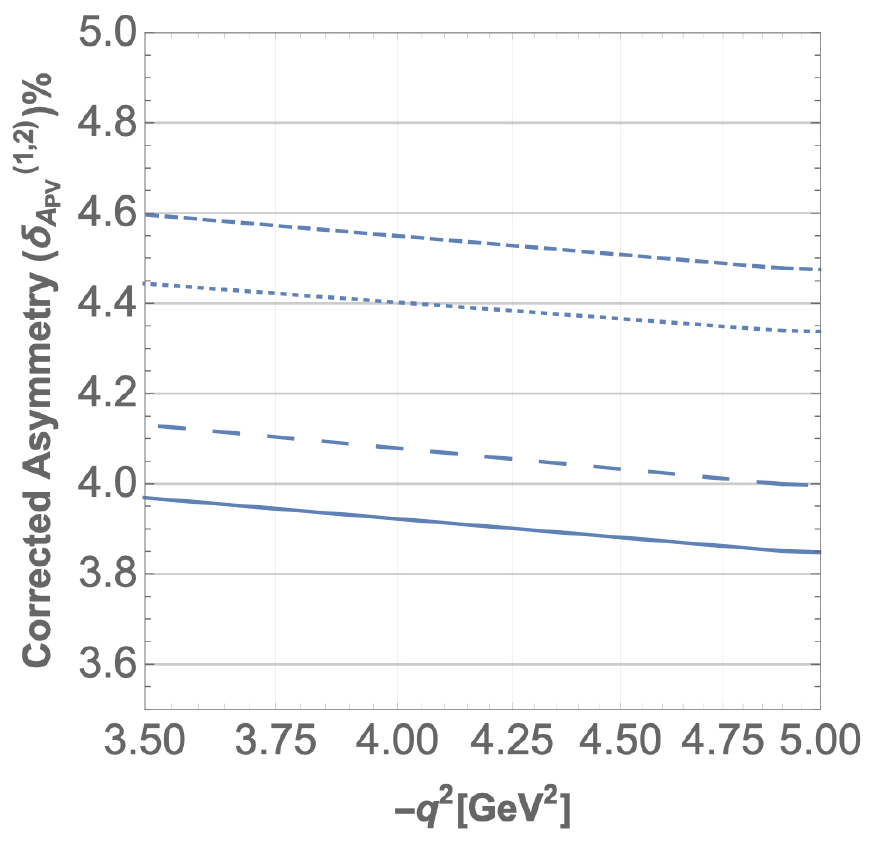}
	\caption{EIC kinematics at $E_{CMS}=20~GeV$: Tree+NLO $\delta^1_{A_{PV}}\%$ (dashed line), Tree+NLO+quadratic $\delta^2_{A_{PV}}\%$ (dotted line), Tree+NLO+two-loop reducible $\delta^2_{A_{PV}}\%$ (large-dashed line) and Tree+NLO+quadratic+two-loop reducible $\delta^2_{A_{PV}}\%$ (solid line) plotted versus $-q^2$. Here $\delta^1_{A_{PV}}\%$ and $\delta^2_{A_{PV}}\%$ are NLO and NNLO $A_{PV}$ correction percentages as given by Eq.[\ref{eq:delta correction per}].}
	\label{fig:eic correction percent}
\end{figure}

The Electron-Ion Collider (EIC) at the Brookhaven laboratory that aims to precisely measure the constituent quarks and gluons of the proton using a polarized electron-proton beam that will be steered into head-on collisions. This experiment is aimed to be conducted at the center-of-mass energies ranging from $20~GeV$ to $100~GeV$ and gradeable to $\sim 140~GeV$ \cite{EIC.exp}.\\
\indent Using EIC kinematics and considering the center-of-mass energy $20~GeV$, we calculated the higher-order radiative corrections in $A_{PV}$ up to NNLO (quadratic and reducible two-loop) via elastic polarized electron scattering with an unpolarized proton target. These higher-order calculations are a good background check in case of the future EIC experiment and provide new constraints on the polarized elastic $ep$ scattering.\\  
\indent The NLO and NNLO $A_{PV}$ corrections, as well as correction percentages $\delta^1_{A_{PV}}\%$ and $\delta^2_{A_{PV}}\%$ for EIC kinematics at $E_{CMS}=20~GeV$ with different scattering angles $(\theta_{lab})$, are given in Tables[\ref{table:eic correction20}]-[\ref{table:eic correction per20}] and shown graphically in Fig.[\ref{fig:eic correction percent}]. These values are calculated in parts per million $(ppm)$.

\begin{table}[h]
	\centering
     \resizebox{\columnwidth}{!}{%
	\begin{tabular}{||c| c| c| c| c|c ||} 
		\hline
		$\theta_{lab}$ & Tree~$A_{PV}(ppm)$  & 1-loop~$A_{PV}(ppm)$ & Qud-$A_{PV}(ppm)$ & 2-loop~red~$A_{PV}(ppm)$ & Total~$A_{PV}(ppm)$ \\ [0.5ex] 
		\hline\hline
		$0.3^0$ & -72.26 & -68.23 & -68.39 & -68.56 & -68.72 \\ 
		$0.4^0$ & -155.92 & -148.29 & -148.57 & -149.00 & -149.29 \\
		$0.5^0$ & -271.46 & -258.96 & -259.38 & -260.22 & -260.66 \\
		$0.6^0$ & -416.71 & -398.05 & -398.62 & -400.04 & -400.66\\
		$0.8^0$ & -789.54 & -754.79 & -755.71 & -758.77 & -759.79\\
		$0.9^0$ & -1014.43 & -969.80 & -970.89 & -975.03 & -976.26\\
		$1.0^0$ & -1263.37 & -1207.67 & -1208.92 & -1214.32 & -1215.77\\
		$1.1^0$ & -1535.19 & -1467.28 & -1468.68 & -1475.51 & -1477.17\\[1ex] 
		\hline
	\end{tabular}%
    }
	\caption{EIC kinematics at $20~GeV$ CMS energy: Our calculated tree, one-loop, quadratic, reducible two-loop and total (quadratic+reducible two-loop) corrected $A_{PV}$ at different scattering angles $\theta_{lab}$. Here (Qud.) represents the quadratic and (2-loop~red.) is for reducible two-loop level corrected $A_{PV}$. These results are calculated in $ppm$.}
	\label{table:eic correction20}
\end{table}

\begin{table}[h]
	\centering
     \resizebox{\columnwidth}{!}{%
	\begin{tabular}{||c| c| c| c| c ||} 
		\hline
		$\theta_{lab}$ & $\delta^1_{A_{PV}}(1~loop)\%$  & $\delta^2_{A_{PV}}(qud)\%$ & $\delta^2_{A_{PV}}(2~loop~red)\%$ & $\delta^2_{A_{PV}}(qud+2~loop~red)\%$ \\ [0.5ex] 
		\hline\hline
		$0.3^0$ & 5.57 & 5.35 & 5.12 & 4.89 \\ 
		$0.4^0$ & 4.89 & 4.71 & 4.44 & 4.25 \\
		$0.5^0$ & 4.61 & 4.45 & 4.14 & 3.98 \\
		$0.6^0$ & 4.48 & 4.34 & 3.99 & 3.85 \\
		$0.8^0$ & 4.40 & 4.28 & 3.89 & 3.77 \\
		$0.9^0$ & 4.39 & 4.29 & 3.88 & 3.76 \\
		$1.0^0$ & 4.41 & 4.31 & 3.88 & 3.77 \\
		$1.1^0$ & 4.42 & 4.33 & 3.89 & 3.78 \\[1ex]
		\hline
	\end{tabular}%
    }
	\caption{EIC kinematics for $20~GeV~CMS~energy$: $\delta^1_{A_{PV}}\%$ and $\delta^2_{A_{PV}}\%$ at different scattering angles $\theta_{lab}$. Here $\delta^1_{A_{PV}}\%$ and $\delta^2_{A_{PV}}\%$ are NLO and NNLO $A_{PV}$ correction percentages as given by Eq.[\ref{eq:delta correction per}], whereas (qud.) represents the quadratic and (2 loop~red.) is for reducible two-loop level correction$\%$.}
	\label{table:eic correction per20}
\end{table}

\subsection{MUSE kinematics}\label{muse results}
The MUon Scattering Experiment (MUSE) is proposed to measure $\mu p$ and $ep$ scattering in the same experiment at the same time. This experiment has the potential to demonstrate whether the $\mu p$ and $ep$ interactions are consistent or different, and whether any difference results from beyond the standard model physics or two-photon exchange.\\
\indent We used the proposed kinematics of the MUSE experiment with beam energy in lab reference frame $E_{lab}=0.156~GeV$ and $0.235~GeV$ and calculated up to NNLO electroweak $A_{PV}$ corrections using elastic polarized $\mu p$ and $ep$ scattering. In the case of $E_{lab}=0.156~GeV$, the MUSE $ep$ scattering results are the same as given in subsec.~\ref{p2 results}, whereas at $E_{lab}=0.235~GeV$, the $A_{PV}$ results up to NNLO are plotted in Fig.[\ref{fig:Museep_asy}] versus different scattering angles $\theta_{lab}$ and momentum transfer $(-q^2)$. The MUSE $\mu p$ scattering $A_{PV}$ results with $E_{lab}=0.156~GeV$ and $0.235~GeV$ are plotted in Figs.[\ref{fig:Muse_asy1}]-[\ref{fig:Muse_asy2}] versus $\theta_{lab}$ and $(-q^2)$. The $e p$ and $\mu p$ scattering $A_{PV}$ correction and correction $\%$ values with $E_{lab}=0.156~GeV$ and $0.235~GeV$ at tree, one-loop, quadratic, two-loop reducible, and total (quadratic + two-loop reducible) with respect to $\theta_{lab}\sim (30^0-50^0)$ proposed by the MUSE collaboration \cite{MUSE:2017dod},~\cite{article} are given in Tables[\ref{table:muse ep correction210}]-[\ref{table:muse mup correction per210}]. 

\begin{table}[!h]
	\centering
     \resizebox{\columnwidth}{!}{%
	\begin{tabular}{||c| c| c| c| c|c ||} 
		\hline
		$\theta_{lab}$ & Tree~$A_{PV}(ppb)$  & 1-loop~$A_{PV}(ppb)$ & Qud-$A_{PV}(ppb)$ & 2-loop~red~$A_{PV}(ppb)$ & Total~$A_{PV}(ppb)$ \\ [0.5ex] 
		\hline\hline
		$30^0$ & -171.27 & -118.89 & -119.99 & -118.36 & -119.46 \\ 
		$35^0$ & -248.25 & -173.98 & -175.45 & -172.98 & -174.44 \\
		$40^0$ & -347.35 & -245.81 & -247.70 & -244.06 & -245.94 \\
		$45^0$ & -473.05 & -338.12 & -340.46 & -335.29 & -337.62\\
		$50^0$ & -630.29 & -455.11 & -457.94 & -450.77 & -453.59\\[1ex] 
		\hline
	\end{tabular}%
    }
	\caption{MUSE kinematics for $ep$ scattering at $E_{lab}=0.235~GeV$: Our calculated tree, one-loop, quadratic, reducible two-loop and total (quadratic+reducible two-loop) $A_{PV}$ at different scattering angles $\theta_{lab}$. Here (Qud.) represents the quadratic and (2-loop~red.) is for reducible two-loop level corrected $A_{PV}$. These results are calculated in $ppb$.}
	\label{table:muse ep correction210}
\end{table}

\begin{table}[!h]
	\centering
     \resizebox{\columnwidth}{!}{%
	\begin{tabular}{||c| c| c| c| c ||} 
		\hline
		$\theta_{lab}$ & $\delta^1_{A_{PV}}(1~loop)\%$  & $\delta^2_{A_{PV}}(qud)\%$ & $\delta^2_{A_{PV}}(2~loop~red)\%$ & $\delta^2_{A_{PV}}(qud+2~loop~red)\%$ \\ [0.5ex] 
		\hline\hline
		$30^0$ & 30.59 & 29.94 & 30.89 & 30.25 \\ 
		$35^0$ & 29.92 & 29.32 & 30.32 & 29.73 \\
		$40^0$ & 29.23 & 28.69 & 29.74 & 29.19 \\
		$45^0$ & 28.52 & 28.03 & 29.12 & 28.63 \\
		$50^0$ & 27.79 & 27.35 & 28.48 & 28.04 \\[1ex] 
		\hline
	\end{tabular}%
    }
	\caption{MUSE ($ep$) scattering kinematics at $E_{lab}=0.235~GeV$: $\delta^1_{A_{PV}}\%$ and $\delta^2_{A_{PV}}\%$ at different scattering angles $\theta_{lab}$. Here $\delta^1_{A_{PV}}\%$ and $\delta^2_{A_{PV}}\%$ are NLO and NNLO $A_{PV}$ correction percentages as given by Eq.[\ref{eq:delta correction per}], whereas (qud.) represents the quadratic and (2 loop~red.) is for reducible two-loop level correction$\%$.}
	\label{table:muse ep correction per210}
\end{table}

\begin{figure}[!htb]
	\centering
	\includegraphics[scale=0.5]{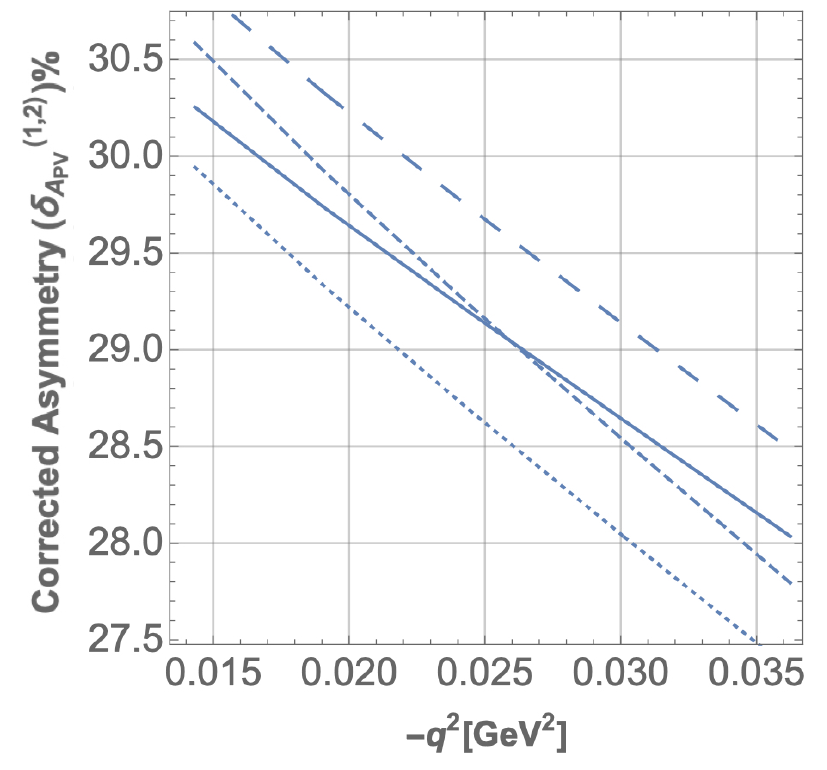}
	\caption{MUSE kinematics for $ep$ scattering at $E_{lab}=0.235~GeV$: Tree+NLO $\delta^1_{A_{PV}}\%$ (dashed line), Tree+NLO+quadratic $\delta^2_{A_{PV}}\%$ (dotted line), Tree+NLO+two-loop reducible $\delta^2_{A_{PV}}\%$ (large-dashed line) and Tree+NLO+quadratic+two-loop reducible $\delta^2_{A_{PV}}\%$ (solid line) plotted versus $-q^2$. Here $\delta^1_{A_{PV}}\%$ and $\delta^2_{A_{PV}}\%$ are NLO and NNLO $A_{PV}$ correction percentages as given by Eq.[\ref{eq:delta correction per}].}
	\label{fig:museep correction percent}
\end{figure}

The MUSE $ep$ scattering $A_{PV}$ correction percentage at $E_{lab}=0.235~GeV$ is shown graphically in Fig.[\ref{fig:museep correction percent}].

\begin{table}[!h]
	\centering
     \resizebox{\columnwidth}{!}{%
	\begin{tabular}{||c| c| c| c| c|c ||} 
		\hline
		$\theta_{lab}$ & Tree~$A_{PV}(ppb)$  & 1-loop~$A_{PV}(ppb)$ & Qud-$A_{PV}(ppb)$ & 2-loop~red~$A_{PV}(ppb)$ & Total~$A_{PV}(ppb)$ \\ [0.5ex] 
		\hline\hline
		$30^0$ & -24.89 & -17.02 & -16.88 & -16.84 & -16.69 \\ 
		$35^0$ & -33.89 & -23.29 & -23.12 & -23.04 & -22.88 \\
		$40^0$ & -44.34 & -30.63 & -30.43 & -30.32 & -30.12 \\
		$45^0$ & -56.29 & -39.07 & -38.83 & -38.68 & -38.44\\
		$50^0$ & -69.82 & -48.68 & -48.38 & -48.21 & -47.91\\[1ex] 
		\hline
	\end{tabular}%
    }
	\caption{MUSE kinematics for $\mu p$ scattering at $E_{lab}=0.156~GeV$: Our calculated tree, one-loop, quadratic, reducible two-loop and total (quadratic+reducible two-loop) corrected $A_{PV}$ at different scattering angles $\theta_{lab}$. Here (Qud.) represents the quadratic and (2-loop~red.) is for reducible two-loop level corrected $A_{PV}$. These results are calculated in $ppb$.}
	\label{table:muse mup correction115}
\end{table}

\begin{table}[!h]
	\centering
     \resizebox{\columnwidth}{!}{%
	\begin{tabular}{||c| c| c| c| c ||} 
		\hline
		$\theta_{lab}$ & $\delta^1_{A_{PV}}(1~loop)\%$  & $\delta^2_{A_{PV}}(qud)\%$ & $\delta^2_{A_{PV}}(2~loop~red)\%$ & $\delta^2_{A_{PV}}(qud+2~loop~red)\%$ \\ [0.5ex] 
		\hline\hline
		$30^0$ & 31.62 & 32.12 & 32.33 & 32.89 \\ 
		$35^0$ & 31.29 & 31.78 & 32.01 & 32.49 \\
		$40^0$ & 30.93 & 31.38 & 31.64 & 32.08 \\
		$45^0$ & 30.59 & 31.03 & 31.29 & 31.72 \\
		$50^0$ & 30.28 & 30.71 & 30.95 & 31.37 \\[1ex] 
		\hline
	\end{tabular}%
    }
	\caption{MUSE kinematics for $E_{lab}=0.156~GeV$: $\delta^1_{A_{PV}}\%$ and $\delta^2_{A_{PV}}\%$ at different scattering angles $\theta_{lab}$. Here $\delta^1_{A_{PV}}\%$ and $\delta^2_{A_{PV}}\%$ are NLO and NNLO $A_{PV}$ correction percentages as given by Eq.[\ref{eq:delta correction per}], whereas (qud.) represents the quadratic and (2 loop~red.) is for reducible two-loop level correction$\%$.}
	\label{table:muse mup correction per115}
\end{table}
\begin{table}[!h]
	\centering
     \resizebox{\columnwidth}{!}{%
	\begin{tabular}{||c| c| c| c| c|c ||} 
		\hline
		$\theta_{lab}$ & Tree~$A_{PV}(ppb)$  & 1-loop~$A_{PV}(ppb)$ & Qud-$A_{PV}(ppb)$ & 2-loop~red~$A_{PV}(ppb)$ & Total~$A_{PV}(ppb)$ \\ [0.5ex] 
		\hline\hline
		$30^0$ & -115.15 & -81.18 & -80.77 & -80.41 & -80.01 \\ 
		$35^0$ & -164.04 & -116.66 & -116.04 & -115.61 & -114.99 \\
		$40^0$ & -225.46 & -161.81 & -160.91 & -160.44 & -159.54 \\
		$45^0$ & -301.61 & -218.54 & -217.29 & -216.78 & -215.54\\
		$50^0$ & -394.93 & -289.00 & -287.34 & -286.79 & -285.14\\[1ex] 
		\hline
	\end{tabular}%
    }
	\caption{MUSE kinematics for $\mu p$ scattering at $E_{lab}=0.235~GeV$: Our calculated tree, one-loop, quadratic, reducible two-loop and total (quadratic+reducible two-loop) corrected $A_{PV}$ at different scattering angles $\theta_{lab}$. Here (Qud.) represents the quadratic and (2-loop~red.) is for reducible two-loop level corrected $A_{PV}$. These results are calculated in $ppb$.}
	\label{table:muse mup correction210}
\end{table}

\begin{figure}[!htb]
	\centering
	\includegraphics[scale=0.33]{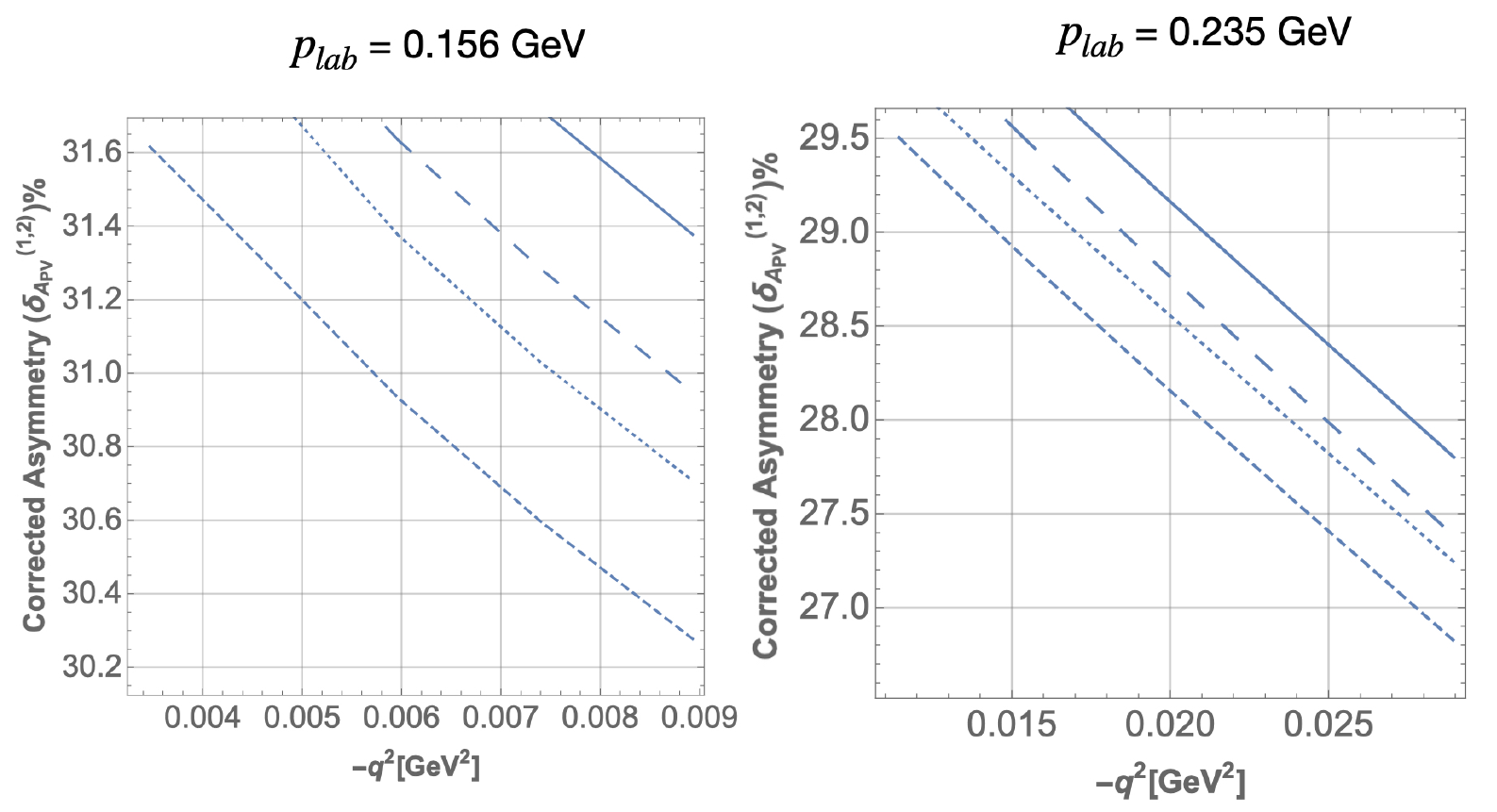}
	\caption{MUSE kinematics for $\mu p$ scattering at $E_{lab}=0.156~GeV$ and $E_{lab}=0.235~GeV$: Tree+NLO $\delta^1_{A_{PV}}\%$ (dashed line), Tree+NLO+quadratic $\delta^2_{A_{PV}}\%$ (dotted line), Tree+NLO+two-loop reducible $\delta^2_{A_{PV}}\%$ (large-dashed line) and Tree+NLO+quadratic+two-loop reducible $\delta^2_{A_{PV}}\%$ (solid line) plotted versus $-q^2$. Here $\delta^1_{A_{PV}}\%$ and $\delta^2_{A_{PV}}\%$ are NLO and NNLO $A_{PV}$ correction percentages as given by Eq.[\ref{eq:delta correction per}].}
	\label{fig:muse correction percent}
\end{figure}

\begin{table}[!h]
	\centering
    \resizebox{\columnwidth}{!}{%
	\begin{tabular}{||c| c| c| c| c ||} 
		\hline
		$\theta_{lab}$ & $\delta^1_{A_{PV}}(1~loop)\%$  & $\delta^2_{A_{PV}}(qud)\%$ & $\delta^2_{A_{PV}}(2~loop~red)\%$ & $\delta^2_{A_{PV}}(qud+2~loop~red)\%$ \\ [0.5ex] 
		\hline\hline
		$30^0$ & 29.50 & 29.85 & 30.17 & 30.52 \\ 
		$35^0$ & 28.88 & 29.26 & 29.52 & 29.89 \\
		$40^0$ & 28.23 & 28.63 & 28.84 & 29.24 \\
		$45^0$ & 27.54 & 27.95 & 28.13 & 28.54 \\
		$50^0$ & 26.82 & 27.24 & 27.38 & 27.80 \\[1ex] 
		\hline
	\end{tabular}%
          }    
	\caption{MUSE kinematics for $\mu p$ scattering at $E_{lab}=0.235~GeV$:  $\delta^1_{A_{PV}}\%$ and $\delta^2_{A_{PV}}\%$ at different scattering angles $\theta_{lab}$. Here $\delta^1_{A_{PV}}\%$ and $\delta^2_{A_{PV}}\%$ are NLO and NNLO $A_{PV}$ correction percentages as given by Eq.[\ref{eq:delta correction per}], whereas (qud.) represents the quadratic and (2 loop~red.) is for reducible two-loop level correction$\%$.}
	\label{table:muse mup correction per210}
\end{table}
The MUSE $\mu p$ scattering $A_{PV}$ correction percentages at $E_{lab}=0.156~GeV$ and $E_{lab}=0.235~GeV$ are shown graphically in Fig.[\ref{fig:muse correction percent}].
  \section{conclusions}
We have developed various computational techniques in this work which have been checked with the higher order QED corrections in $A_{PV}$. Our analytical and programming routines show considerable promise for extension and applications towards the experiments which are searching for physics beyond the Standard Model.\\
\indent Our numerical results up to NNLO (quadratic + reducible two-loop) level electroweak $A_{PV}$ are in good agreement with the measured values by $Q_{weak}$ experiment as well as predictions for the proposed future P2 experiment. The $Q_{weak}$ measured value of $A_{PV}$ at $\theta_{lab}=7.9^0$ is $-226.5\pm 7.3 (statistical)\pm 5.8(systematic)~ppb$ \cite{Qweak:2013zxf}, whereas our calculated  $A_{PV}$ up to NNLO is $-221.46~ppb$. Similarly, the P2 predicted value of $A_{PV}~$ at $\theta_{lab}=35^0~$ is $-67.34~ppb~$ \cite{P2.exp}, while our calculated $A_{PV}$ at the same $\theta_{lab}$ is $-65.09~ppb.$\\
\indent The theoretical predictions we make in the case of polarized $e p$ and $\mu p$ scattering using the kinematics of highly anticipated experiments MOLLER and MUSE, and programs at EIC will play an important role either directly or in the background studies for the search of physics beyond the Standard Model at the precision frontier. These calculations can further be improved by adding electroweak box diagrams. Since box diagrams are self-gauge invariant, we can calculate them separately, and so our next target is to investigate different approaches for the calculation of boxes. In the future, we would also like to further extend this work by considering a polarized proton target in the case of both elastic and inelastic scattering scenarios. This will be achieved by calculating a polarized hadronic tensor at higher momentum transfer.
\begin{acknowledgments}
 We would like to acknowledge the financial support from the Natural Sciences and Engineering Research Council of Canada (NSERC) and the CFI. M. Ghaffar, would also like to thank MUN for graduate scholarship and CINP for the travel support during this project.
\end{acknowledgments}

\appendix

\section{\label{leptonic structure functions appendix}LEADING ORDER ELECTROWEAK LEPTONIC TENSOR STRUCTURE FUNCTIONS}
The Leading Order (LO) electroweak leptonic tensor structure functions with a polarized incoming electron having a helicity state $\beta=\pm1$ are given below:
\begin{multline*}
l_1=\frac{\alpha \pi}{4c^3_Ws^3_W}\Bigg(4q^2c^2_Ws^2_W(-1+4s^2_W)+\\
c_Ws_W(-2m^2_l+q^2+4q^2s^2_W)+2\beta m_l(c_Ws_W\\
-4s^2_W(c^2_W+c_Ws_W))(s_1\cdot k_2)\Bigg),\\
\\
l_2=\frac{\alpha \pi}{2c^2_Ws^2_W}\Bigg(1+4c_Ws_W(2c_Ws_W+4s^2_W-1)-4s^2_W(1-2s^2_W)\Bigg),\\
\\
l_3=\frac{\alpha \pi}{2c^2_Ws^2_W}\Bigg(1+4c_Ws_W(2c_Ws_W+4s^2_W-1)-4s^2_W(1-2s^2_W)\Bigg),\\
\\
l_4=\frac{2\beta \alpha m_l\pi}{c^3_Ws_W}\Bigg(c_Ws_W(2s^2_W-1)+c^2_W(2c_Ws_W+4s^2_W-1)\Bigg),\\
\\
l_5=-\frac{\beta \alpha m_l\pi}{2c^3_Ws^3_W}\Bigg(4c^2_Ws^2_W(4s^2_W-1)+c_Ws_W(1+4s^2_W)\Bigg),\\
\\
l_6=\frac{\alpha \pi}{2c^2_Ws^2_W}\Bigg(1-4s_W(c_W+s_W)\Bigg),\\
\\
l_7=\frac{\beta \alpha m_l\pi}{2c^2_Ws^2_W}\Bigg(4s_W(c_W+s_W)-1\Bigg),\\
\\
l_8=\frac{\beta \alpha m_l\pi}{2c^2_Ws^2_W}\Bigg(4s_W(c_W+s_W)-1\Bigg).
\end{multline*}
\section{\label{hadronic structure functions appendix}LEADING ORDER ELECTROWEAK HADRONIC TENSOR STRUCTURE FUNCTIONS}
The LO unpolarized electroweak hadronic tensor structure functions are given as:
\begin{multline*}
H^{VV}_1=\frac{\alpha \pi C_2 G^2_A }{4c^2_Ws^2_W}\Bigg(F^{\gamma}_{2n}+(2+F^{\gamma}_{2p})(4s_W(c_W+s_W)-1)\Bigg),\\
\\
H^{VV}_2=-\frac{\alpha \pi C_2}{128m^2_pc^2_Ws^2_W}\Bigg((F^{\gamma}_{2n})^2(q^2+4m^2_p)+\\
2F^{\gamma}_{2n}(4s_W(c_W+s_W)-1)(4m^2_p(2+F^{\gamma}_{2p})+q^2F^{\gamma}_{2p})+\\
F^{\gamma}_{2p}(1+8s^2_W+8c_Ws_W(4s^2_W-1))(4m^2_p(4+F^{\gamma}_{2p})+q^2F^{\gamma}_{2p})\Bigg),\\
\\
H^{VV}_3=\frac{\alpha \pi C_2}{128m^2_pc^2_W s^2_W}\Bigg(4m^2_p(8+(F^{\gamma}_{2n})^2+4F^{\gamma}_{2p}+(F^{\gamma}_{2p})^2+\\
8G^2_A-8c_Ws_W(8+4F^{\gamma}_{2p}+(F^{\gamma}_{2p})^2)+8(8+F^{\gamma}_{2p}(4+F^{\gamma}_{2p}))\\
\times(2c_W(c_W+2s_W)-1)s^2_W+16(8+F^{\gamma}_{2p}(4+F^{\gamma}_{2p}))s^2_W+\\
2F^{\gamma}_{2n}(2+F^{\gamma}_{2p})(4s_W(c_W+s_W)-1))-(q^2(F^{\gamma}_{2n}-F^{\gamma}_{2p})(F^{\gamma}_{2n}+\\
F^{\gamma}_{2p}(8c_Ws_W-1))+8F^{\gamma}_{2p}(F^{\gamma}_{2n}+F^{\gamma}_{2p}(2c_W(c_W+\\
2s_W)-1))s^2_W+16(F^{\gamma}_{2p})^2s^2_W)\Bigg),\\
\\
H^{VV}_4=-\frac{\alpha \pi C_2}{128 m^2_pc_Ws_W}\Bigg((F^{\gamma}_{2n})^2(q^2+4m^2_p)+\\
2F^{\gamma}_{2n}(4s_W(c_W+s_W)-1)(4m^2_p(2+F^{\gamma}_{2p})+q^2F^{\gamma}_{2p})+\\
F^{\gamma}_{2p}(1+8s^2_W+8c_Ws_W(4s^2_W-1))(4m^2_p(4+F^{\gamma}_{2p})+
q^2F^{\gamma}_{2p}\Bigg),\\
\\
H^{VV}_5=\frac{\alpha \pi C_2}{128m^2_pc^2_W s^2_W}\Bigg(4m^2_p(8+(F^{\gamma}_{2n})^2+4F^{\gamma}_{2p}+(F^{\gamma}_{2p})^2+\\
8G^2_A-8c_Ws_W(8+4F^{\gamma}_{2p}+(F^{\gamma}_{2p})^2)+
8(8+F^{\gamma}_{2p}(4+F^{\gamma}_{2p}))\\
\times(2c_W(c_W+2s_W)-1)s^2_W+
16(8+F^{\gamma}_{2p}(4+F^{\gamma}_{2p}))s^2_W+\\
2F^{\gamma}_{2n}(2+F^{\gamma}_{2p})(4s_W(c_W+s_W)-1))-(q^2(F^{\gamma}_{2n}-F^{\gamma}_{2p})(F^{\gamma}_{2n}+\\
F^{\gamma}_{2p}(8c_Ws_W-1))+
8F^{\gamma}_{2p}(F^{\gamma}_{2n}+F^{\gamma}_{2p}(-1+\\
2c_W(c_W+2s_W)))s^2_W+16(F^{\gamma}_{2p})^2s^2_W)\Bigg),\\
\\
H^{VV}_6=\frac{\alpha \pi C_2}{32c^2_Ws^2_W}\Bigg(q^2((F^{\gamma}_{2n})^2+2F^{\gamma}_{2n}(2+F^{\gamma}_{2p})\\
\times(4s_W(c_W+s_W)-1)+(2+F^{\gamma}_{2p})^2(1+8s^2_W+\\
8c_Ws_W(4s^2_W-1)))+4G^2_A(q^2-4m^2_p)\Bigg).\\
\\
\end{multline*}

Here the terms $F^{\gamma}_{(1,2)p}$ and $F^{\gamma}_{(1,2)n}$ are the electric and magnetic form factors of the proton and neutron, respectively. The values of these form factors at $q^2\rightarrow 0$ are given by:
\begin{equation*}
\begin{split}
F^{\gamma}_{1p}(q^2=0)=1,\\
F^{\gamma}_{2p}(q^2=0)=2.793-1,\\
F^{\gamma}_{1n}(q^2=0)=0,\\
F^{\gamma}_{2n}(q^2=0)=-1.9147.
\end{split}
\end{equation*}
The structure of the form factors is represented by the term $C_i\equiv \left(\frac{\Lambda^2}{\Lambda^2-q^2}\right)^i$ with $\Lambda=\sqrt{0.83}m_p$. For simplicity, we used the dipole structure of the form factors given by $C_2\equiv \left(\frac{\Lambda^2}{\Lambda^2-q^2}\right)^2 $.\\
\indent For higher order (NLO and NNLO) electroweak leptonic and hadronic tensor structure functions are too cumbersome to show here and we can provide the analytical expressions at the request from the authors.
\section{\label{Bremsstrahlung appendix}ONE AND TWO PHOTON EXCHANGE SOFT BREMSSTRAHLUNG}
 The soft photon integral term $I(k_i,k_j)$ has the following form:
 \begin{equation}
	I(k_i,k_j)=\frac{4\pi \alpha}{(2\pi)^3}\Bigg(-2m^2_lk_{1,1}+(2m^2_l-q^2)k_{1,2}\Bigg),
\label{eq:spbintegral}
\end{equation}
where $k_{i,j}$ terms in Eq.[\ref{eq:spbintegral}] are defined as:
\begin{multline*}
	k_{i,j}=\frac{2\pi l_{i,j}}{l_{i,j}^2m^2_{i}-m^2_{j}}
	\Bigg(\frac{1}{2}
	\text{ln}\left[\frac{l_{i,j}^2m^2_{j}}{m^2_{j}}\right]\text{ln}\left[\frac{4~\Delta E^2}{\lambda}\right]+\\
	\frac{1}{4}\left(\text{ln}\left[\frac{E_{i}-p}{E_{i}+p}\right]\right)^2-\frac{1}{4}\left(\text{ln}\left[\frac{E_{j}-p}{E_{j}+p}\right]\right)^2+\\
	\text{Li}_2\left[2,1-\frac{l_{i,j}}{v_{i,j}}(E_{i}+p)\right]+
	\text{Li}_2\left[2,1-\frac{l_{i,j}}{v_{i,j}}(E_{i}-p)\right]-\\
	\text{Li}_2\left[2,1-\frac{1}{v_{i,j}}(E_{j}+p)\right]-
	\text{Li}_2\left[2,1-\frac{1}{v_{i,j}}(E_{j}-p)\right]\Bigg),
\end{multline*}
where Li$_2$ represents the Spence function or dilogarithm whose properties are given in \cite{THOOFT1979365}. The rest of the terms are defined as:
  \begin{equation*}
  \begin{split}
   	m^2_{1,2}\equiv m^2_l,\\
    \\
		E_{1,2} =\sqrt{p^2+m^2_l},\\
        \\
		p^2 = {\frac{(E_{CMS}^2+m^2_p-m^2_l)}{4E_{CMS}^2}-m^2_p},\\
        \\
		v_{i,j} = \frac{l_{i,j}^2m^2_{i}-m^2_{j}}{2(l_{i,j}E_{i}-E_{j})},\\
        \\
		l_{1,1}= 1,\\
        \\
		l_{1,2} = 1-\frac{q^2}{2m^2_l}+\frac{\sqrt{q^4-4q^2 m^2_l}}{2m^2_l}.\\
        \end{split}
     \end{equation*}
The terms $E_{1,2}$ are the energies of the incoming and outgoing leptons. The term $p^2$ is the spatial momentum squared for the incoming and outgoing leptons in the center of mass reference frame with $E_{CMS}$ as the center-of-mass energy. The terms $m_l$ and $m_p$ represent the lepton and proton mass, respectively.

\bibliography{MG-AA-SB-EW}

\end{document}